\documentclass[a4paper,11pt]{article}

\usepackage{jcappub} % for details on the use of the package, please
\usepackage[T1]{fontenc} % if needed
\usepackage{multirow}
\usepackage[table]{xcolor}
\usepackage{tabularx}

\newcommand{\dd}{\text{d}}

\usepackage{siunitx} % for physics units
\usepackage{booktabs} % To thicken table lines (toprule, bottomrule...)

\title{Setting requirements on out-of-band rejection for next-generation CMB experiments. \newline \Large{Application to the \textit{LiteBIRD} instrument}}
\author[1]{L.\,Mousset,}
\author[2]{L.\,Montier,}
\author[2]{J.\,Aumont,}
\author[10,3,4]{F.\,Columbro,}
\author[3,4]{P.\,de~Bernardis,}
\author[5]{J.\,Errard,}
\author[6, 7]{C.\,Franceschet,}
\author[8]{S.\,Giardiello,}
\author[9, 10]{T.\,Ghigna,}
\author[11]{H.\,Hubmayr,}
\author[11]{G.\,Jaehnig,}
\author[3,4]{S.\,Masi,}
\author[3, 4]{F.\,Piacentini,}
\author[3]{G.\,Pisano,}
\author[12]{A.\,Rizzieri,}
\author[13]{G.\,Savini,}
\author[8]{C.\,Tucker,}

\author[ ]{\\LiteBIRD Collaboration.}

\affiliation[1]{Université Paris-Saclay, CNRS, Institut d’Astrophysique Spatiale, 91405, Orsay, France}
\affiliation[2]{IRAP, Université de Toulouse, CNRS, CNES, UPS, Toulouse, France}
\affiliation[3]{Dipartimento di Fisica, Università La Sapienza, P. le A. Moro 2, Roma, Italy}
\affiliation[4]{INFN Sezione di Roma, P.le A. Moro 2, 00185 Roma, Italy}
\affiliation[5]{Université Paris Cité, CNRS, Astroparticule et Cosmologie, F-75013 Paris, France}
\affiliation[6]{Dipartimento di Fisica, Università degli Studi di Milano, Via Celoria 16 - 20133, Milano, Italy}
\affiliation[7]{INFN Sezione di Milano, Via Celoria 16 - 20133, Milano, Italy}
\affiliation[8]{School of Physics and Astronomy, Cardiff University, Cardiff CF24 3AA, UK}
\affiliation[9]{International Center for Quantum-field Measurement Systems for Studies of the Universe and Particles (QUP), High Energy Accelerator Research Organization (KEK), Tsukuba, Ibaraki 305-0801, Japan}
\affiliation[10]{Kavli Institute for the Physics and Mathematics of the Universe (Kavli IPMU, WPI), UTIAS, The University of Tokyo, Kashiwa, Chiba 277-8583, Japan}
\affiliation[11]{NIST Quantum Sensors Group, 325 Broadway, Boulder, CO 80305, USA}
\affiliation[12]{Department of Physics, University of Oxford, Denys Wilkinson Building, Keble Road, Oxford OX13RH, UK}
\affiliation[13]{Astrophysics Group, Physics and Astronomy Dept, UCL, Gower Street WC1E6BT, London, UK}

\emailAdd{louise.mousset@universite-paris-saclay.fr}

\abstract{Next-generation cosmic microwave background experiments have very stringent constraints to achieve the required sensitivity to target polarization $B$ modes. In this work, we intend to set requirements on the out-of-band rejection level, with out-of-band referring to frequencies outside the telescope band-pass. The method developed is applied to the \textit{LiteBIRD} Medium and High Frequency Telescopes. In order to determine the impact of out-of-band power, we model the instrument's optical response and the spectral emissions of the sky and of the instrument itself. This allows us to propagate optical power inside the telescope. Using this tool, we address both the impact of out-of-band power on the detection chain and on the thermal heat load, together with the impact on the process of separation between astrophysical components. The role of additional static power as well as dynamic power variations is studied. The requirement derived consist in attenuation factors (in dB) in frequency subdomains. They will be used to design the telescopes filters.}

\begin{document}
\maketitle
\flushbottom

%%%%%%%%%%%%%%%%%%%%%%%%%
%%%%%%%%%%%%%%%%%%%%%%%%% 
\section{Introduction}
%%%%%%%%%%%%%%%%%%%%%%%%%
%%%%%%%%%%%%%%%%%%%%%%%%% 

% General context: CMB Bmodes, inflation
Our current understanding of the primordial Universe relies on the cosmological inflation theory~\cite{Brout:1977ix, Starobinsky:1980te, Guth:1980zm}. This phase of accelerated expansion, which occurred at the very beginning of the Universe, is thought to have generated gravitational waves propagating through the primordial plasma, leaving a signature in the polarization anisotropies of the cosmic microwave background (CMB)~\cite{Seljak:1996gy, Kamionkowski:1996ks}. This signature, known as $B$ modes, corresponds to large-scale rotational patterns in the polarization of the CMB. It is characterized by an angular power spectrum whose amplitude is directly proportional to the tensor-scalar ratio (hereafter denoted $r$), which is related to the energy scale at which inflation occurred~\cite{Lyth:1996im}.

% Current experiments, challenging measurement
The current and next-generation CMB experiments are targeting the measurement of the CMB $B$~mode signal. Detecting them would allow us to test the main inflationary models and gain direct insight into the inflationary energy scale. The current upper limit on the tensor-scalar ratio is $r < 0.032$ at 95\% confidence level~\cite{tristramImprovedLimitsTensortoscalar2022}, and it is more and more challenging to improve on this constraint, both for instrumental and data analysis reasons. While currently operating experiments, such as Simons Observatory~\cite{SO_2018}, BICEP-Keck~\cite{BICEP:2021xfz}, ACT~\cite{ACT:2023kun} and SPT~\cite{SPT:2018whf} have already been designed to minimize the impact of systematic effects, the next generation of CMB experiments, from ground and space, has to be designed to achieve both a high sensitivity and an unprecedented control of the instrumental systematic effects to fulfil its scientific objective: measuring $r$ with a precision of $\delta r < 0.001$.

% We need filters from ground and space
Even if the context of CMB observations from ground and space may appear quite different, both have to face some similar obstacles to reach the expected sensitivity on $r$. Ground-based experiments are strongly impacted by the emission from the atmosphere, which imposes a high and unstable load on detectors. Contamination from astrophysical foregrounds is also present, but it is limited by the fact that ground-based experiments generally focus on small areas of the sky where astrophysical foreground emission is relatively weak. On the contrary, space-based experiments benefit from a much more stable environment due to the absence of the atmosphere, however they target the nearly full sky and observe in a wide frequency range, meaning they need to deal with strong contaminations from a variety of complex foreground emission. Thus it appears mandatory in both cases to efficiently make sure that the atmospheric or astrophysical signal out of the observing telescope bands is properly rejected. The objective of this study is to derive requirements for this out-of-band rejection, which will be used to build the filtering strategy for the \textit{LiteBIRD} mission. The modelling described in this article is also part of the preparation for the instrument's calibration tests, which will take place prior to launch, similar to those conducted for the Planck satellite~\cite{Ade_2010_Planck_prelaunch}.

% Plan du papier
The article is organized as follow. In section~\ref{sec:global} we first present the objectives of this study and the methodology that we propose to set out-of-band rejection requirements with particular emphasis on how we build the assumed sky and the instrument models. In section~\ref{sec:LiteBIRD}, we apply this methodology to the \textit{LiteBIRD} instrument case. We first present the status of the mission and the specificities of this instrument, then we derive the requirements from several instrumental constraints (detector chain, thermal cooling budget and thermal stability). We conclude section~\ref{sec:LiteBIRD} with a summary of the requirements we find, pointing out the most stringent ones. Finally, we conclude the paper with a general summary in section~\ref{sec:conclusion}.

%%%%%%%%%%%%%%%%%%%%%%%%%%%%%%%%%%%%%
%%%%%%%%%%%%%%%%%%%%%%%%%%%%%%%%%%%%% 
\section{A global approach to set out-of-band rejection requirements}
%%%%%%%%%%%%%%%%%%%%%%%%%%%%%%%%%%%%%
%%%%%%%%%%%%%%%%%%%%%%%%%%%%%%%%%%%%% 
\label{sec:global}

%%%%%%%%%%%%%%%%%%%%%%%%%%%%%%%%%%%%%%%%%%%%%%%%%%%%%%
\subsection{Objectives}
%%%%%%%%%%%%%%%%%%%%%%%%%%%%%%%%%%%%%%%%%%%%%%%%%%%%%%
\label{ss:obj}

The goal of this section is to present a methodology to put requirements on the knowledge of the spectral response of the detection chain of any instrument operating in the millimetre wave-range. More specifically, we are interested in the characterisation of the out-of-band rejection level associated to the filtering scheme. 

%, which defines the efficiency of power rejection in the spectral range outside the nominal band-pass for a given observing band. 

\begin{figure}[ht!]
    \centering
    \includegraphics[width=1\linewidth]{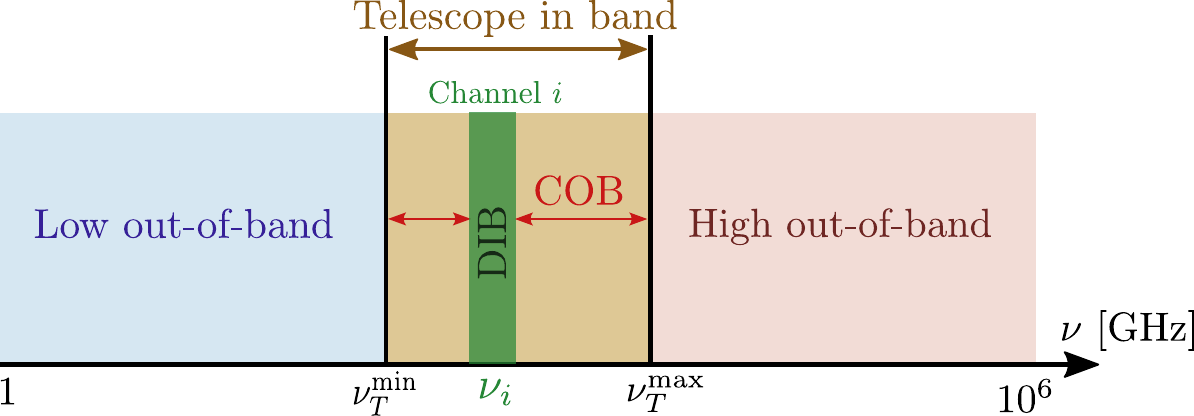}
    \caption{Sketch to illustrate the five frequency domains that we consider. The telescope in-band (TIB) domain, shown in yellow, is split in detector in-band (DIB), shown in green and centred at frequency $\nu_i$, and complementary out-of-band (COB). In blue, we show the low out-of-band (LOB) domain, which extends from 1 GHz to the telescope's lowest frequency $\nu_{T}^{\rm min}$. In red, we show the high out-of-band (HOB) domain, which extends from the telescope's highest frequency $\nu_{T}^{\rm max}$ to $10^6$\,GHz.}
    \label{fig:band_sketch}
\end{figure}

% Frequency domains definition
\label{ss:band_def}
In this study, we consider a total frequency interval spanning from 1 to $10^6$\,GHz, in which we define for each telescope channel, $i$, spanning from $\nu_{i,min}$ to $\nu_{i,max}$, the following frequency domains, related to the admission range of the telescope, as illustrated in figure~\ref{fig:band_sketch}:
\begin{itemize}
    \setlength\itemsep{0em}
    \item telescope in-band (TIB): frequency range covering all observing bands of a single instrument composed by a set of detectors with specific observing bands,
    \item detector in-band (DIB): ideal band-pass top-hat frequency range centred at $\nu_i$ associated to a given detector,
    \item low out-of-band (LOB): frequency range below the TIB frequency range,  
    \item high out-of-band (HOB): frequency range above the TIB frequency range, 
    \item complementary out-of-band  (COB): frequencies included in the TIB frequency range, but outside the DIB range.
\end{itemize}
In the following, we refer to these domains with either TIB, DIB, LOB, COB, HOB, or with T, D, L, C, H, indiscriminately. 

For a given frequency channel $i$, the total power $P_i$ per detector, can be written as: 
\begin{equation}
P_i = \sum_{c} P_{c,i} = \sum_{c} \int_{\nu} P_{c}^{\rm \, in} (\nu) \cdot H_{i} (\nu) d\nu \, ,
\end{equation}
where $P_{c,i} $ is the power received by the detector from the component $c$, $P_{c}^{\rm \,in}(\nu)$ is the incident power spectral density from component $c$, and $H_{i}(\nu)$ is the transfer function of the channel $i$. By component, we mean here any component from the astrophysical sky or elements of the instrument itself. 

For a given frequency domain $d \in [L, C, D, H]$, we restrict the frequency range and we have: 
\begin{equation}
P_i^d = \sum_{c} \int_{\nu^{d}_{min}}^{\nu^{d}_{max}} P^{\, \rm in}_{c} (\nu) \cdot H_{i} (\nu) d\nu \,.
\end{equation}
Then, we assume that in each out-of-band frequency domain the instrument transfer function can be decomposed into a frequency dependent part, $H_{i}^{inst}$ and a constant term, $A_{i}^{d}$
\begin{equation}
\forall \nu \in [\nu_{min}^{d},\nu_{max}^{d}]\, , \,  H_{i}(\nu) = H_{i}^{inst}(\nu) \cdot A_{i}^{d}\, .
\end{equation}
$A_{i}^{d}$ takes values $\leq 1$, and it is to be interpreted as the attenuation introduced by the filtering scheme.
Hence the total power received by the detector from the channel $i$ can be written as
\begin{align}
P_i &= \sum_{d} A_{i}^{d} \left( \sum_{c} \int_{\nu_{min}^{d}}^{\nu_{max}^{d}} P_{c}^{\rm \, in} (\nu) \cdot H^{inst}_{i} (\nu) d\nu \right)\\
&= \sum_{d} A_{i}^{d} \sum_c \widetilde P_{c,i}^{\,d} \\
&= \sum_{d} A_{i}^{d} \widetilde P_i^{\,d} \quad \text{for} \quad d \in[L,C,D,H] \,, \label{eq:total_power}
\end{align}
where we have introduced $\widetilde P_i$ ($\widetilde P_i^{\,d}, \widetilde P_{c,i}^{\,d}$) to indicate total power (per domain and per component) that would be received in the absence of filters.
Actually, the attenuation coefficients, $A_{i}^{d}$, defined at the detector level are obtained by a sequence of filters positioned at various locations of the optical path. We can thus rewrite $A_{i}^{d}$ as a product of the attenuation coefficients of each of these filters:
\begin{equation}
A_{i}^{d} = \prod_s A_{i,s}^{d} \, ,
\end{equation}
where $A_{i,s}^{d}$ is the attenuation coefficient of the channel $i$ inside the frequency domain $d$ at the position $s$ of the optical path.  

We assume that $A_{i,s}^D=1$, i.e. the detector band-pass inside the band is assumed to be ideal. The goal of this work is to put requirements on this set of $A_{i,s}^d$ quantities for $d=L,C,H$
to minimize the impact of the out-of-band power on the performances of the instrument. These recovered requirement values can then be injected as input specifications to design a telescope filtering scheme.

%%%%%%%%%%%%%%%%%%%%%%%%%%%%%%%%%%%%%%%%%%%%%%%%%%%%%%
\subsection{Methodology}
%%%%%%%%%%%%%%%%%%%%%%%%%%%%%%%%%%%%%%%%%%%%%%%%%%%%%%
\label{ss:methodo}

In order to put requirements on the out-of-band attenuation coefficients, $A_{i,s}^{L,C,H}$, we would in principle start from a set of high-level requirements on science and on hardware, that we would flow down through the instrumental response in all observing frequency bands.

In the case of CMB $B$~mode experiments, such as \textit{LiteBIRD}, the high-level scientific requirement is generally defined by the total uncertainty $\delta r$ obtained for the tensor-to-scalar ratio. On the instrumental side, the high-level requirement is associated to the thermal constraints of the instrument, both in terms of available cooling power to cool down the optics and the focal plane, and in terms of temperature stability requirements at the various cryogenic stages. 

In practice, to achieve this flow-down process, we start from a set of assumptions on the out-of-band attenuation coefficients, $A_{i,s}^{L,C,H}$, and propagate their impact through the whole detection chain up to the reconstruction of the $r$ parameter and the hardware high-level properties. This process requires to assume both a detailed modelling of the instrumental response without any filtering scheme (see section~\ref{ss:instru_model}), and a modelling of the astrophysical sky (see section~\ref{ss:skymodel}), in order to compute the $\widetilde P_{c,i}^{\,d}$ quantities and reconstruct the total power received by a detector using Eq.~\ref{eq:total_power} (see section~\ref{sss:power_det_generic}). These input power quantities can then be propagated up to the sensitivity-per-band performances (see section~\ref{sss:sensitvt_computation}) and instrumental thermal performances (see section~\ref{sss:radiative_load}). We then include in the analysis a component separation method, which allows us to combine the signal from all the bands to extract the foreground-cleaned CMB maps and assess the impact on the sensitivity on $r$. We finally compare the outcomes of this forecast analysis with the high-level requirements of the experiment in order to accept or discard the initial assumed $A_{i,s}^{L,C,H}$ values. Repeating the procedure for different assumed $A_{i,s}^{L,C,H}$ values, we find the requirements on $A_{i,s}^{L,C,H}$ as the largest (then less stringent) values compliant with high-level objectives.  

The approach followed in this work includes all major aspects of the instrument when exploring the impact of the out-of-band rejection levels. We address both the impact on the detection chain and on the thermal heat load, which are key ingredients of the instrument performances, together with the impact on the process of separation between astrophysical components when combining multiple observing bands to finally extract the science from the CMB signal. Hence we consider as input parameters of this analysis the possible deviations of the incident power due to out-of-band signal coming from the emission of the observed sky ($P_{\rm sky}^{L,C,H}$) and from the instrument itself ($P_{\rm inst}^{L,C,H}$), those deviations being assumed to be static ($\Delta P$) or dynamic ($\delta P$), as summarised in table~\ref{tab:method}. We thus perform two sub-analysis, referred to as static and dynamic analysis.

In the case of the static analysis, we first propagate the effect of the out-of-band power coming from the sky and the instrument to the detection chain by assessing the impact on the detectors noise equivalent power (NEP) (see section~\ref{sss:NEP}), and on the component separation process in section~\ref{sss:comp_sep}. We then address the impact of the out-of-band power on the thermal balance of the instrument (see section~\ref{ss:req_from_thermal}), to assess the compliance with the cooling power capability, both at the focal plane, $P^{\, \rm cool}_{\rm{det}}$, and optical element levels, $P^{\, \rm cool}_{\rm{inst}}$.

For the dynamic analysis, we propagate the effect of the fluctuations of the out-of-band power coming from the sky signal on the detection chain by assessing the impact on the detector NEP (see section~\ref{sss:dynamic_NEP}). We also propagate the sky and instrument temperature fluctuations up to the detector bath temperature (see section~\ref{sss:dynamic_Tdet}) and to the various optical elements (see section~\ref{sss:dynamic_Tinst}), in order to check the compliance with the expected temperature stability  of these cryogenic stages, i.e. $\delta T_{\rm{inst}}$ and $\delta T_{\rm{det}}$.

By exploring the various configurations described above, we can build a complete set of requirements to be applied on the $A_{i,s}^{L,C,H}$ coefficients, at various positions $s$ on the optical path, for each frequency channel $i$ of the instrument.

\begin{table}
\center

%\begin{tabular}{| l | m{1.5cm} | m{1.5cm} | m{1.5cm} | m{1.5cm} |}
\begin{tabular}{
      | l |
      >{\centering}m{1.5cm} |
      >{\centering}m{1.5cm} |
      >{\centering}m{1.5cm} |
      >{\centering\arraybackslash}m{1.5cm} |
      }
\hline
input perturbation & \multicolumn{4}{c|}{assessment criteria} \\
\hline
& \multicolumn{2}{c|}{detection chain} & \multicolumn{2}{c|}{thermal balance} \\
\hline
\hline
\textbf{Static} & \centering $\Delta$NEP & \centering Comp Sep &  \centering $P^{\,\rm cool}_{\rm inst}$ &  $P^{\, \rm cool}_{\rm det}$
\\
\hline
$\Delta P_{\rm{sky}}^{L,C,H}$ & 
\ref{sss:NEP} & 
\ref{sss:comp_sep} & 
\ref{sss:HWP_heat_load} \ref{sss:lenses_heat_load} & 
\ref{sss:FP_heat_load} \\
\hline
$\Delta P_{\rm{inst}}^{L,C,H}$ &  
\centering \ref{sss:NEP} & 
\cellcolor{gray}  & 
\cellcolor{gray} & 
\ref{sss:FP_heat_load} \\ 
\hline
\hline
\textbf{Dynamic} & \multicolumn{2}{c|}{$\Delta$NEP}  & {\centering  $\delta T_{\rm inst}$} &  {\centering $\delta T_{\rm det}$}  \\ 
\hline
$\delta P_{\rm{sky}}^{L,C,H}$ & 
\multicolumn{2}{c|}{\ref{sss:dynamic_NEP}}  &
\ref{sss:dynamic_Tinst} & 
\ref{sss:dynamic_Tdet}\\
\hline
$\delta P_{\rm{inst}}^{L,C,H}$ &  
\multicolumn{2}{c|}{\ref{sss:dynamic_NEP}} &
\cellcolor{gray}  & 
 \ref{sss:dynamic_Tdet} \\
\hline
\end{tabular}
 
\caption{Overview of the various configurations studied to assess the out-of-band rejection levels. For each input perturbation due to the presence of out-of-band power, the impact is estimated in regards to an assessment criteria and described in the corresponding section.}
\label{tab:method}
\end{table}

%%%%%%%%%%%%%%%%%%%%%%%%%%%%%%%%%%%%%%%
\subsection{Modelling the emissivity of the Sky}
%%%%%%%%%%%%%%%%%%%%%%%%%%%%%%%%%%%%%%%
\label{ss:skymodel}

\begin{figure}
    \centering
    \includegraphics[width=1\linewidth]{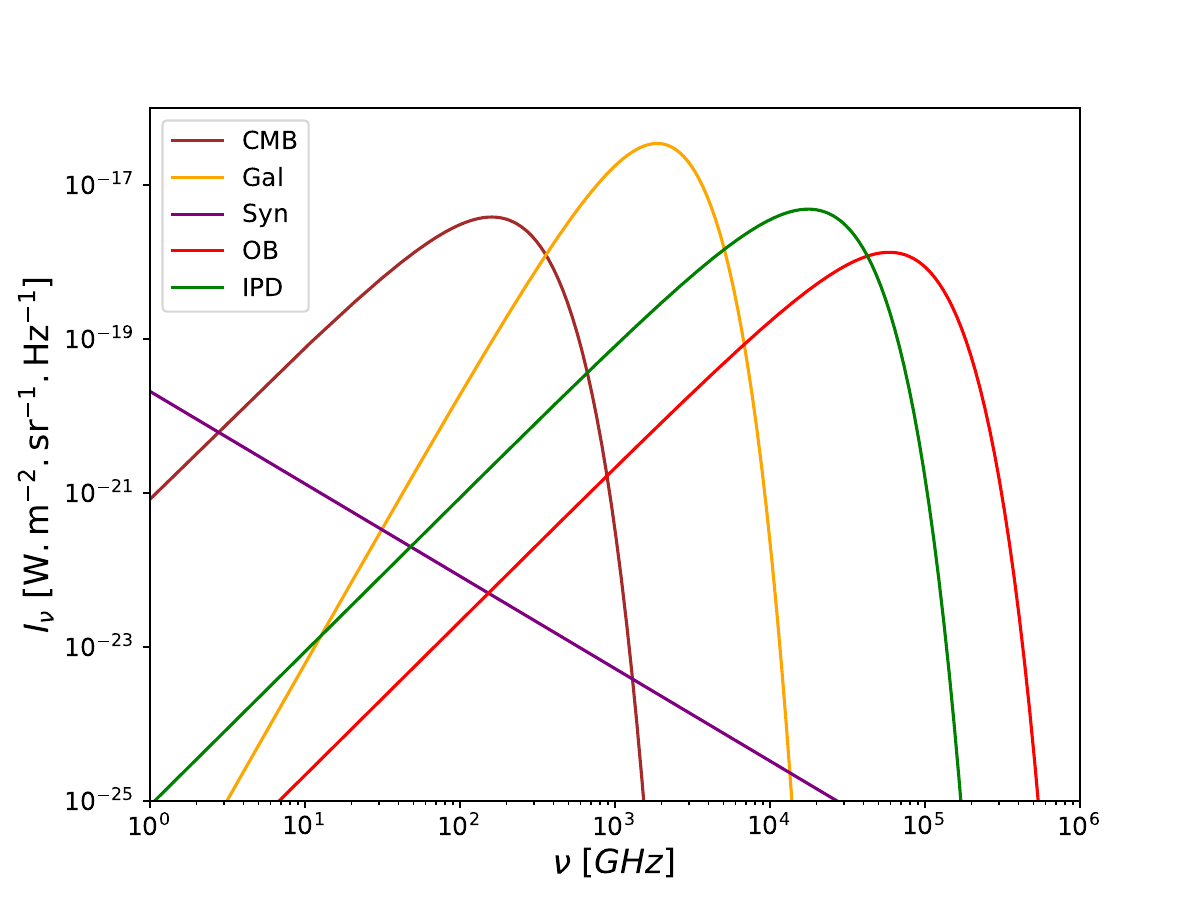}
    \caption{Spectral radiance of the five sky components: CMB, Galactic dust, synchrotron, interplanetary dust (IPD), the most brilliant stars (O and B types). The CMB is modelled as a black-body at 2.725\,K. For the other components, the spectral radiance is scaled on measurements assuming a one degree FWHM beam size (generic example).}
    \label{fig:skyspectra}
\end{figure}

We consider emissions from five astrophysical components: CMB, thermal Galactic dust, synchrotron, zodiacal light due to interplanetary dust (IPD) and the emission from the most brilliant stars (O and B~types). While the CMB radiation is mostly isotropic, the Galactic dust, synchrotron and stars are brighter on the Galactic plane. On the contrary, the IPD concentrates in the ecliptic plane. In this section we detail the assumptions made to build the spectral radiance of each component, shown in figure~\ref{fig:skyspectra}. We consider a frequency range from 1 to $10^6$\,GHz which adequately covers the spectral range of the sky components under consideration. This is particularly true at high frequencies, which are the most concerning in terms of contamination.

The CMB is modelled as a black-body with temperature $T_0=\SI{2.725}{\kelvin}$~\cite{Fixsen:2009ug}. Its spectral radiance is given by the Planck radiation law:
\begin{equation}
    \label{eq:BB_CMB}
    I_{\rm CMB}(\nu) = \frac{2 h \nu^3}{c^2} \frac{1}{e^{\frac{h \nu}{k T_0}}-1} 
    \quad [\rm{W}.\rm{m}^{-2}.\rm{sr}^{-1}.\rm{Hz}^{-1}] \,,  
\end{equation}
where $\nu$ is the frequency, $h$ is the Planck constant, $c$ the speed of light, and $k$ the Boltzmann constant. 

The Galactic dust Spectral Energy Density (SED) is modelled as a modified black-body~\cite{Hensley:2017ygd}:
\begin{equation}
    \label{eq:BBmodified_dust}
    I_{\rm dust}(\nu) \propto B_{\nu}(T_{\rm{d}}) \left(\frac{\nu}{\nu_d}\right)^{\beta_d}
    \quad [\rm{MJy}.\rm{sr}^{-1}]
\end{equation}
with $B_{\nu}(T)$ the Planck radiation law and $\nu_d=353$\,GHz the reference frequency~\cite{Planck:2013ltf}. Moreover, we have used the d0 model from the Python Sky Model (\texttt{PySM}) library~\cite{Thorne:2016ifb} which assumes a fixed spectral index of $\beta_d = 1.54$ and a black-body temperature of $T_d = \SI{20}{\kelvin}$. 

Concerning the synchrotron emission, the SED is modelled as a power law with spectral index $\beta_s = - 1.2$~\cite{Hensley:2017ygd}:
\begin{equation}
    \label{eq:synchrotron}
    I_{\rm syn}(\nu) \propto \left(\frac{\nu}{\nu_s} \right)^{\beta_s}
    \quad [\rm{MJy}.\rm{sr}^{-1}].
\end{equation}
The reference frequency is $\nu_s = \SI{408}{\mega\hertz} $ in order to follow the s0 \texttt{PySM} template in intensity~\cite{Remazeilles:2014mba}. 

The IPD emission is modelled by a grey-body at $T_{\rm IPD} = 300$\,K with a very small emissivity~\cite{Staubach1993TemperaturesOZ, Takimoto_2022} such that
\begin{equation}
    \label{eq:BBmodified_IPD}
    I_{\rm IPD}(\nu) \propto B_{\nu}(T_{\rm IPD}) 
    \quad [\rm{MJy}.\rm{sr}^{-1}]
\end{equation}

Similarly, the emission from OB stars is modelled as a grey-body at $T_{\rm OB} = 1000$\,K:
\begin{equation}
    \label{eq:BBmodified_OB}
    I_{\rm OB}(\nu) \propto B_{\nu}(T_{\rm OB}) 
    \quad [\rm{MJy}.\rm{sr}^{-1}]
\end{equation}

%Amplitudes of the SED
The SED amplitudes of Galactic dust, synchrotron radiation, the IPD, and OB stars are scaled against available measurements. For each component, we use template maps measured at reference frequencies. Since the objective of this article is to define requirements for the filtering scheme, we take a conservative approach. For each component, we set the intensity at the reference frequency to the maximum value of the template map. In this way, the incident sky power entering the instrument corresponds to its expected maximum value. 

For Galactic dust and synchrotron, we have used the template maps given by \texttt{PySM} models s0 and d0. For IPD we used the map from the Cosmic Background Explorer Diffuse Infrared Brightness Experiment (COBE/DIRBE) at \SI{25}{\micro\meter}~\cite{Kelsall:1998bq, Planck:2013ltf} and for OB stars, we have considered the map from the WISE mission at \SI{12}{\micro\meter}~\cite{WISE2014}.

Moreover, the maximum intensity depends on the spatial resolution of the map. Therefore, before taking the maximal pixel value, the map is smoothed at the instrument beam resolution. When considering a detector from channel $i$ the map is smoothed at that channel resolution. We end up with a reference intensity, and so a spectral radiance, that depends on the frequency channel $i$. As an example, in figure~\ref{fig:skyspectra}, the SED of the five sky components are shown considering a spatial resolution of one degree.

%Polarization
Finally, the five components described here are partially polarized. The CMB is linearly polarized at a 10\,\% level but considering only the $B$ mode polarization, it is expected to be about $10^3$ smaller than temperature anisotropies. For Galactic dust and synchrotron emissions, we have considered 10\,\% and 48\,\% polarisation fractions respectively, as modelled in~\cite{Hensley:2017ygd}. The IPD polarization fraction is taken to be 1\,\% which is the current limit from ISO/CAM measurements in the mid-infrared~\cite{Ganga:2021xhc}. Finally, stars are very little polarized, their polarization fraction is set to $10^{-4}$~\cite{Cotton_2016}.

%%%%%%%%%%%%%%%%%%%%%%%%%%%%%%%%%%%%%%%%%%%%%%%%%%%%%%%%
\subsection{Instrument modelling}
%%%%%%%%%%%%%%%%%%%%%%%%%%%%%%%%%%%%%%%%%%%%%%%%%%%%%%%%
\label{ss:instru_model}

In the following we restrict the study to a refractive optical design. This allows us to write explicitly the power received by a detector, called $\widetilde P_i^d$, as a function of the optical element properties.

%------------------------------------------------------
\subsubsection{Simple refractive design}
%------------------------------------------------------
\label{sss:refractive_design}

\begin{figure}
    \centering
    \includegraphics[width=1\linewidth]{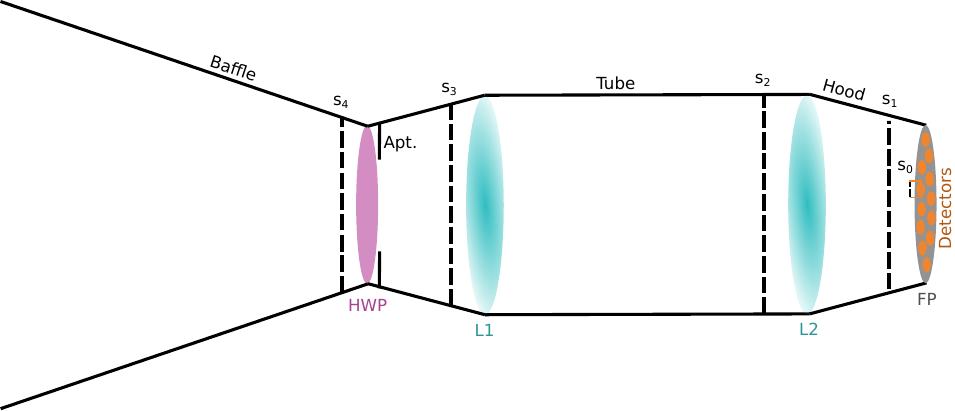}
    \caption{Refractive optics modelling. We represent the mechanical structure (baffle, aperture stop, tube and hood), the HWP, the two lenses (L1 and L2) and the focal plane paved with detectors. Possible filter positions $s_0$ to $s_4$ are indicated with dash lines. $s_0$ is a on-chip filter.}
    \label{fig:optical_sketch}
\end{figure}

We consider a very general refractive telescope design, sketched in figure~\ref{fig:optical_sketch}. The instrument is composed of a baffle mounted on a tube which contains two lenses L1 and L2 that focus the signal on a focal plane (FP) paved with detectors. We also consider a half-wave plate (HWP) at the entrance of the tube. The tube between L2 and the FP is called the hood.

Each mechanical or optical element is assumed to have a homogeneous mean temperature. The temperature will typically decrease along the optical path to reach a minimum value on the focal plane, thanks to cryogenic stages which are not specified in this model. In our study, the spectral radiance of any element is simply modelled by the black body law at the mean temperature. 

Optical properties are described in terms of emissivity $E(\nu)$, reflectivity $R(\nu)$ and efficiency $\varepsilon(\nu)$ (or transmission) which are functions of the frequency. Those quantities are related through:
\begin{equation}
    \label{eq:emissivity}
    E(\nu) + R(\nu) + \varepsilon(\nu) = 1\, .
\end{equation}

As described in section~\ref{ss:obj}, the total attenuation coefficient $A_i^d$ at detector level can be decomposed in a product of sub-attenuation coefficients $A_{i,s}^d$ where $s$ corresponds to the coordinate along the optical path. Under this design assumption, we can specify the possible location of filters, shown as dash lines in figure~\ref{fig:optical_sketch}. We have considered five positions: in front of the HWP, between the HWP and L1, between L1 and L2, between L2 and the FP, as well as on-chip filters associated to a single detector.

%------------------------------------------------------
\subsubsection{Power received by a detector}
%------------------------------------------------------
\label{sss:power_det_generic}

Under this refractive design assumption, we can explicitly write the optical power received by a detector, defined in section~\ref{ss:obj} as $\widetilde P_i^d$. This power corresponds to the received power assuming no filters, i.e. $A_{i, s}^d=1 \, \forall \, (i, s, d)$. 

We consider the five sky components, as described in section~\ref{ss:skymodel}, and the power emitted by the instrument itself, from the optical elements (HWP, L1 and L2) and from the mechanical structure (baffle, tube and hood). Then, the power in Watt falling on a detector from a given component $c$, coming from one of the fourth frequency domains $d$, for one detector channel $i$, can be written as
\begin{equation}
    \label{eq:power-det}
    \widetilde P^d_i = S \Omega_i \sum_c \int_{\nu^d_{\rm min}}^{\nu^d_{\rm max}} E_c(\nu) \varepsilon_{c \rightarrow \rm{det}}(\nu) I_{c, (i)}(\nu) \dd\nu \, ,
\end{equation}
where $S$ is the collector area, taken as the HWP surface, $\Omega_i$ is the solid angle for the channel $i$ and $I_{c, (i)}(\nu)$ is the spectral radiance of the component $c$ in $\si{\watt \per \square\meter  \per \steradian \per \hertz}$. The latter for sky components are modelled as presented in section~\ref{ss:skymodel}. In that case they depend on the detector channel $i$, because of the way the sky SED are normalized: reference intensity equal to the maximum of a map smoothed at the channel $i$ resolution. For the instrument sub-systems instead, spectral radiances are modelled as black-bodies at their mean temperature and in that case they do not depend on the channel, hence the parenthesis around $i$. $E_c (\nu)$ is the emissivity of component $c$, taken to be 1 for sky components and which has to be modelled for each instrument component. Finally, $\varepsilon_{c \rightarrow \rm{det}} (\nu)$ is the total efficiency of optical elements along the optical path between the emitting source for component $c$ and the detector. The total efficiency is the product of individual efficiencies of the optical elements along the light path, including the factor of two due to single polarization receiver.

%------------------------------------------------------
\subsubsection{Sensitivity computation}
%------------------------------------------------------
\label{sss:sensitvt_computation}

To asses the instrument performance, we need to estimate the channel sensitivities corresponding to a given design. This is for instance the purpose of \texttt{Bolocalc}~\cite{Hill:2018rva} software developed for Simons Observatory.

Knowing the incident power on detectors, one can compute the corresponding photon NEP. Then, assuming a model for other sources of NEP (for example readout electronic, thermal fluctuations, mechanical vibrations...), one can compute the total detector NEP. The NEP can be converted in CMB noise equivalent temperature (NET) expressed in \si{\kelvin_{CMB}\sqrt\second}. Finally, from the detector NET, one can compute the total NET of the full focal plane array and the corresponding channel sensitivity in \si{\kelvin_{CMB}}-arcmin with some assumptions on the scanning strategy (observing time and observed sky fraction in particular).

%------------------------------------------------------
\subsubsection{Radiative load on optical elements}
%------------------------------------------------------
\label{sss:radiative_load}

In a similar manner as we did for the computation of the received power at detector level, we can express the radiative load on optical elements. From the spectral radiances of each emitting component $c_1$, we compute the radiative heat load on the HWP, L1, L2 or the FP. The power received by an optical component $c_2$ in the frequency domain $d$ can be written as:
\begin{equation}
    \label{eq:power-thermal}
    \widetilde P^d_{c_2} = \sum_{c_1} \left ( S_{c_2}\Omega_{c_1\rightarrow c_2}
    \int_{\nu^d_{\rm{min}}}^{\nu^d_{\rm{max}}} E_{c_1}(\nu) \varepsilon_{c_1 \rightarrow c_2}(\nu) E_{c_2}(\nu) I_{c_1}(\nu) \dd\nu \right )\, .
\end{equation}
The term $S_{c_2}\Omega_{c_1\rightarrow c_2}$ is the optical extent, $S_{c_2}$ is the collector area of component $c_2$ and $\Omega_{c_1\rightarrow c_2}$ the solid angle under which $c_2$ sees $c_1$. The spectral radiance $I_{c_1}(\nu)$ is multiplied by the emissivity $E_{c_1}(\nu)$ of the emitting component, the total efficiency $\varepsilon_{c_1 \rightarrow c_2}(\nu)$ of optical elements along the optical path, and the absorption of the receiver which, at thermal equilibrium, is equal to its emissivity so named $E_{c_2}(\nu)$. As before, when $c_1$ is a sky component, the emissivity is set to 1 and the sky spectral radiance is normalized by a reference intensity. In that case, since we are interested in the integrated flux entering into the instrument and illuminating the full focal plane, the reference intensity is extracted from a measured map smoothed at the resolution of the full field of view of the telescope.

%%%%%%%%%%%%%%%%%%%%%%%%%%%%%%%%%%%%%%%%%%%%
%%%%%%%%%%%%%%%%%%%%%%%%%%%%%%%%%%%%%%%%%%%%
\section{Application to the \textit{LiteBIRD} case}
%%%%%%%%%%%%%%%%%%%%%%%%%%%%%%%%%%%%%%%%%%%%
%%%%%%%%%%%%%%%%%%%%%%%%%%%%%%%%%%%%%%%%%%%%
\label{sec:LiteBIRD}
While the methodology described above is very general, in the following we have applied it to the \textit{LiteBIRD} instrument. After describing the \textit{LiteBIRD} design specificities in section~\ref{ss:LB_instru}, we will derive requirements on the filtering scheme following section~\ref{ss:methodo}. This is the purpose of sections~\ref{ss:req_from_det}, \ref{ss:req_from_thermal} and~\ref{ss:req_from_dynamic}. Finally we conclude with a summary of all the requirements derived in section~\ref{ss:req_summary}.

%%%%%%%%%%%%%%%%%%%%%%%%%%%%%%%%%%%%%%%%%%%%
\subsection{The \textit{LiteBIRD} instrument}
%%%%%%%%%%%%%%%%%%%%%%%%%%%%%%%%%%%%%%%%%%%%
\label{ss:LB_instru}

%------------------------------------------------------
\subsubsection{Project overview}
%------------------------------------------------------
\label{sss:LB_overview}
\textit{LiteBIRD} is a JAXA-led Strategic Large-Class mission, selected by ISAS/JAXA in 2019 to be launched in the 2030s, and aimed at mapping the CMB polarized emission over the full sky at large angular scales. \textit{LiteBIRD} has been endorsed as one of the prioritized projects in the Master Plan 2020 of the Science Council of Japan. The project is currently supported in Phase A by many other partners, including France, Italy, UK, Germany, Spain, and Canada. While the concept design has been studied by researchers from Japan, the U.S., Canada, and Europe since September 2016, it has recently been improved under the supervision of space agencies. More specifically, since September 2024, a new effort of simplification of the design and procurement has been undertaken, with the help of the whole collaboration. This process led to the replacement of the previous three-telescope configuration with a single instrument and a single detection chain. The primary scientific objective has been modified to achieve a total uncertainty in the tensor-to-scalar ratio of $\delta r < 0.002$ (68\% C.L.), keeping in mind that this now includes contributions from foreground residuals, statistical uncertainties, instrumental systematics, and margin contingency, while detail budget has not been defined yet~\cite{LiteBIRD:2026gdf}.

However, in the context of this work we will still rely on the former design, which is based on three distinct telescopes, as described in Sec.~\ref{sss:LB_design}. More generally, we will use the same set of assumptions as used in the reference paper~\cite[][hereafter PTEP]{PTEP}, which presents the global specifications of the \textit{LiteBIRD} instruments and high-level requirements. The goal here is to illustrate the methodology and, in the future, we plan to adapt this study to the new instrument configuration. In fact, this work is already underway within the \textit{LiteBIRD} collaboration.

%------------------------------------------------------
\subsubsection{Telescope design}
%------------------------------------------------------
\label{sss:LB_design}

Following the PTEP paper, the \textit{LiteBIRD} instrument proposed design is divided in three telescopes spanning the frequency range from 40 to 448\,GHz, they are called Low-, Medium and High-Frequency Telescopes (respectively LFT, MFT and HFT). In addition of the PTEP paper, a detailed presentation of MFT and HFT can be found in~\cite{LiteBIRD:2020zfx}. MFT and HFT both share the same optical and mechanical designs. They are refractive telescopes with two plastic lenses with assumed index of refraction $n_r = 1.52$. The diffraction-limited field of view is $28\deg$. LFT is instead a reflective telescope with mirrors.

The MFT and HFT telescopes are expected to cover the frequency range 89–448\,GHz, split over 10~frequency channels. Table~\ref{tab:channels} gives the main parameters for each frequency channel $i$. Detector beams are well approximated by a Gaussian, with mean equal to the central frequency of each channel and full width at half maximum (FWHM) evaluated through ray tracing simulations~[PTEP].

\begin{table}
    \centering
    \begin{tabular}{|c|ccccc|ccccc|}
        \hline
        & \multicolumn{5}{c|}{\textbf{MFT}} & \multicolumn{5}{c|}{\textbf{HFT}} \\
        \hline
         \textbf{Channel [GHz]} & 100 & 119 & 140 & 166 & 195 & 195 & 235 & 280 & 337 & 402 \\
         \textbf{Bandwidth} & 0.23 & \multicolumn{4}{c|}{0.3} & \multicolumn{4}{c}{0.3} & 0.23 \\
         \textbf{FWHM [arcmin]} & 37.8 & 33.6 & 30.8 & 28.9 & 28.0 & 28.6 & 24.7 & 22.5 & 20.9 & 17.9 \\
         \textbf{Number of pixels} & 183 & 244 & 183 & 244 & 183 & 127 & 127 & 127 & 127 & 169\\
         \hline
    \end{tabular}
    \caption{MFT and HFT frequency channel parameters: central frequency, bandwidth, beam FWHM and the number of pixels. Each pixel on the focal plane contains two detectors sensitive to orthogonal polarizations.}
    \label{tab:channels}
\end{table}

Both MFT and HFT focal-plane units use superconducting Transition-Edge Sensors (TES) bolometers~\cite{Jaehnig:2020dun} cooled down to 100\,mK using a system based on seven adiabatic demagnetization refrigerator (ADR) stages that allow a 100\,\% duty cycle~\cite{Duval:2020rit}. The focal planes are composed of 3428 detectors (2074 for MFT and 1354 for HFT). The TES bolometers are coupled with silicon lenslets and sinuous antennas for MFT and coupled with silicon platelet feedhorns and orthomode-transducer feeds for HFT. The focal planes include monochromatic, dichroic and trichroic pixels which are sensitive to polarization~\cite{Suzuki:2018cuy}.

Reaching the scientific goal on $r$ requires a high sensitivity and strong mitigation of the $1/f$ noise. This led the \textit{LiteBIRD} collaboration to consider for all telescopes the use of a continuously rotating HWP placed at the entrance of each telescope~[PTEP]. The presence of this rotating HWP performs an effective suppression of the $1/f$ noise, allowing us to distinguish between the instrumental polarized signal and the sky signal, which is modulated at four times the angular speed of the HWP. The HWP temperature is maintained at approximately 20\,K.

%------------------------------------------------------
\subsubsection{Optical and mechanical modelling}
%------------------------------------------------------
\label{sss:LB_assumption}

%%%% IMO_Perf
This work as been done with a Python software developed within the \textit{LiteBIRD} collaboration in order to model the instrument and estimate its performance. This code is called \texttt{IMo\_Perf} for Instrument Model Performance and follows the general method described in section~\ref{sss:sensitvt_computation}. Thus, the main objective of this code is to compute the sensitivity of each channel, knowing the incident power on the detectors. Additional features and details are given in Appendix~\ref{app:NEP}. Today, \texttt{IMo\_Perf} is the reference code used by the collaboration to calculate the instrument's sensitivities. We list here the assumptions made for the instrument mechanical and optical properties.

\begin{figure}
    \centering
    \includegraphics[width=1\linewidth]{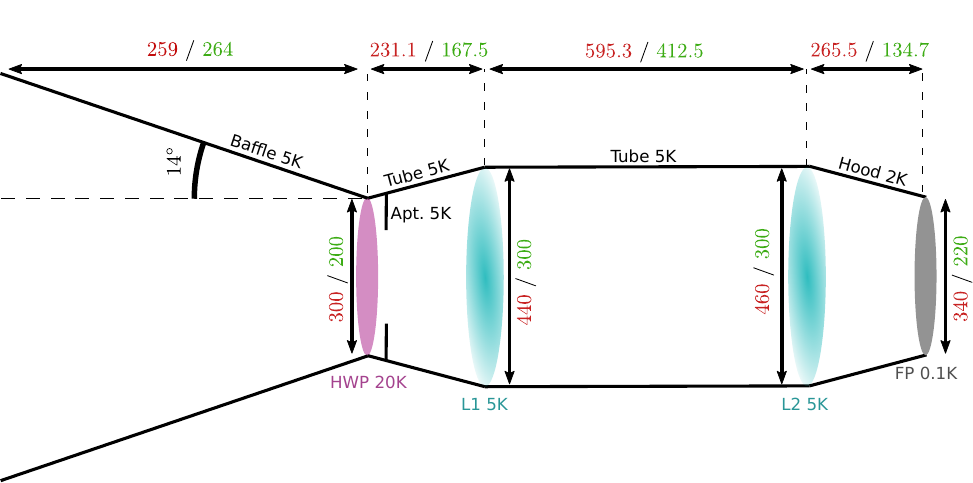}
    \caption{Optical sketch of the MFT/HFT. Optical and mechanical elements are named and we report their temperatures. The dimensions in millimetres are written for both instruments, MFT in red and HFT in green.}
    \label{fig:MHFT_sketch}
\end{figure}

%%%% Mechanical geometry
While the general refractive design studied in this paper was presented in section~\ref{ss:instru_model}, figure~\ref{fig:MHFT_sketch} shows the specificities of MFT and HFT in terms of dimensions and temperature at the different stages. The cryogenic chain is not modelled, we only assume that the instrument has reached a stationary and nominal thermal behaviour.  

%%%% HWP behaviour
In this model we do not consider any instrumental systematic effect. In particular, the HWP behaves ideally and does not induce any cross-polarization. The modulation efficiency is modelled by a fraction between 0.95 and 0.99 depending on the frequency channel, based on optical modelling. This represents the fraction of polarized signal that is effectively modulated at four times the HWP rotation speed. 

%%%% Emissivité 
\begin{figure}
    \centering
    \includegraphics[width=0.8\linewidth]{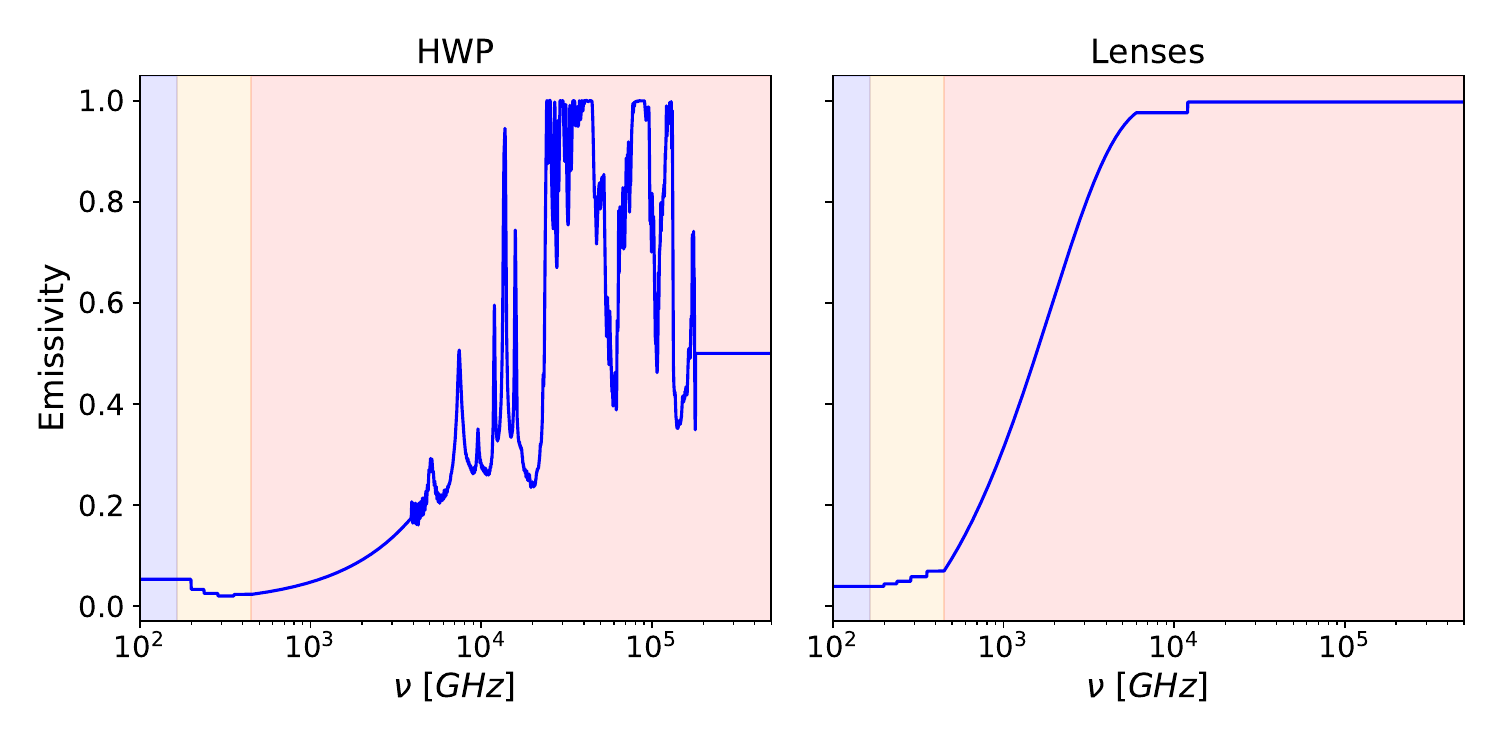}
    \caption{Model of the optical emissivity as a function of frequency for the HWP and the lenses for the HFT instrument. The background colors correspond to the frequency ranges: LOB in blue, TIB in yellow, and HOB in red.}
    \label{fig:Emiss_OOB}
\end{figure}

% In-band
Regarding the optical properties of the HWP and lenses (emissivity, efficiency, reflectivity) those quantities are known for each detector channel, based on optical modelling.\footnote{The HWP model was built using the high frequency structure simulation (HFSS) software: \url{https://www.ansys.com/products/electronics/ansys-hfss}. The lenses were modelled with the Zemax software.} They are considered constant within each frequency channel $i$. When it is not necessary to make an explicit distinction between frequency channels -- for example, in section~\ref{ss:req_from_thermal} -- we calculate the average across all channels.

% OoB
For out-of-band frequencies, we built a model based on data measured in the laboratory. The HWP is made of polypropylene (PP) and the lenses are made of polyethylene (PE). We know the transmission of PP between approximately 3.9\,THz and 180\,THz (see figure 6 in~\cite{Tucker_2006}), which partially covers the HOB range. This measurement was made for a PP thickness of 1\,mm, which is a good order of magnitude for the HWP. Above 180\,THz, as there is no data available, we arbitrarily set the emission to 0.5. Between the maximal detector frequency channel and 3.9\,THz, we simply chose to use a linear model. For the LOB domain, we took a constant emissivity value, equal to that of the minimum frequency detector channel. This emissivity model is shown in Figure~\ref{fig:Emiss_OOB} on the left.

For the lenses, made of PE, we have similar transmission measurements available within the \textit{LiteBIRD} collaboration. This measurement was performed for a PE thickness of approximately 2.3 mm, between 1.5\,THz and 18\,THz, at a temperature of 1.5\,K. From this measurement, we extracted a rough emissivity profile: equal to 0.5 above 12\,THz, equal to 0.35 between 6 and 12\,THz, and linear between the maximum detector channel value and 6\,THz. Finally, since the lenses are approximately 2\,cm thick and not 2.3\,mm, we had to rescale the emissivity knowing that the transmission can be written as:
\begin{equation}
    T(\nu) = \exp{\left(-\alpha(\nu) e\right)}
\end{equation}
where $\alpha$ is the absorption coefficient and $e$ is the thickness. Finally, for the LOB domain, we proceed as for the HWP, we take a constant emissivity value, equal to that of the minimum frequency detector channel. The emissivity model for the lenses, covering the full frequency range, is shown in Figure~\ref{fig:Emiss_OOB} on the right.

Concerning the mechanical structure, baffle, tube and hood, in order to limit reflections inside the instrument they are covered with an absorber. Their emissivity is thus taken to be 1 at any frequency. 

%%% Solid angle
The solid angle is computed assuming a Gaussian beam, $\Omega_i = \frac{\pi \theta_i^2}{4 \ln 2}$ where $\theta_i$ is the $\text{FWHM}$ of channel $i$ given in table~\ref{tab:channels}. This expression is correct in the detector band, as the optics is adjusted to be at the diffraction limit in the nominal frequency range. For out-of-band frequencies, this hypothesis is conservative as the optical coupling with detectors is not optimal so that the effective optical extent is smaller.

% Optical extent
In section~\ref{ss:req_from_thermal}, we compute the thermal heat load on the HWP, L1, L2 and FP. For this purpose, we need to estimate the optical extent associated with each pair of optical elements (HWP-L1, L1-L2,\dots) radiating on each other. Solid angles are estimated, assuming a simple geometrical model. The method and the hypotheses that were made are presented in appendix~\ref{app:solid_angle}.

%------------------------------------------------------
\subsubsection{Budget error allocation}
%------------------------------------------------------
\label{sss:budget_error}
The main scientific driver for the \textit{LiteBIRD} mission is to achieve a total uncertainty on $r$ measurement such as $\delta r < 0.001$. This requirement includes contributions from instrumental statistical noise fluctuations,  instrumental systematics, residual foregrounds, lensing $B$ modes, and observer bias, and shall not rely on future external data sets~[PTEP].
The total budget assigned for systematic error is $6.5 \times 10^{-4}$ for the three telescopes and the 22 frequency channels. For individual systematics we assign 1\% of the total budget to account for possible dozens of systematics, i.e. $6.5 \times 10^{-6}$.

%%%%%%%%%%%%%%%%%%%%%%%%%%%%%%%%%%%%%%%%%%%%%%%%%%%%%%%%%%%%%
\subsection{Requirements from detection chain - Static analysis}
%%%%%%%%%%%%%%%%%%%%%%%%%%%%%%%%%%%%%%%%%%%%%%%%%%%%%%%%%%%%%
\label{ss:req_from_det}

The goal of this section is to derive constraints on out-of-band filtering from detector chain properties. Since in this section we deal with incident power on a single detector, we intend to put constraints on the attenuation factor in the LOB, COB and HOB domains. The factors that we constrain are the total ones that combine all positions, that is, $A_i^d = \prod_s A_{i,s}^d$ for $d=L, C, H$.

In section~\ref{sss:power_det}, we quantify the increase of the incident power on a detector due to the out-of-band contribution assuming no filtering. Then, we address two main effects due to out-of-band power: the increase of the detector NEP (see section~\ref{sss:NEP}), and the modification of component separation efficiency (see section~\ref{sss:comp_sep}). Finally, since this study was conducted assuming an ideal HWP, in section~\ref{sss:HWP_margin} we justify this assumption by looking at the impact of a non-ideal HWP.

%------------------------------------------------------
\subsubsection{Incident power on a detector assuming no filtering}
%------------------------------------------------------
\label{sss:power_det}

\begin{figure}[ht!]
    \centering
    \includegraphics[width=1\linewidth]{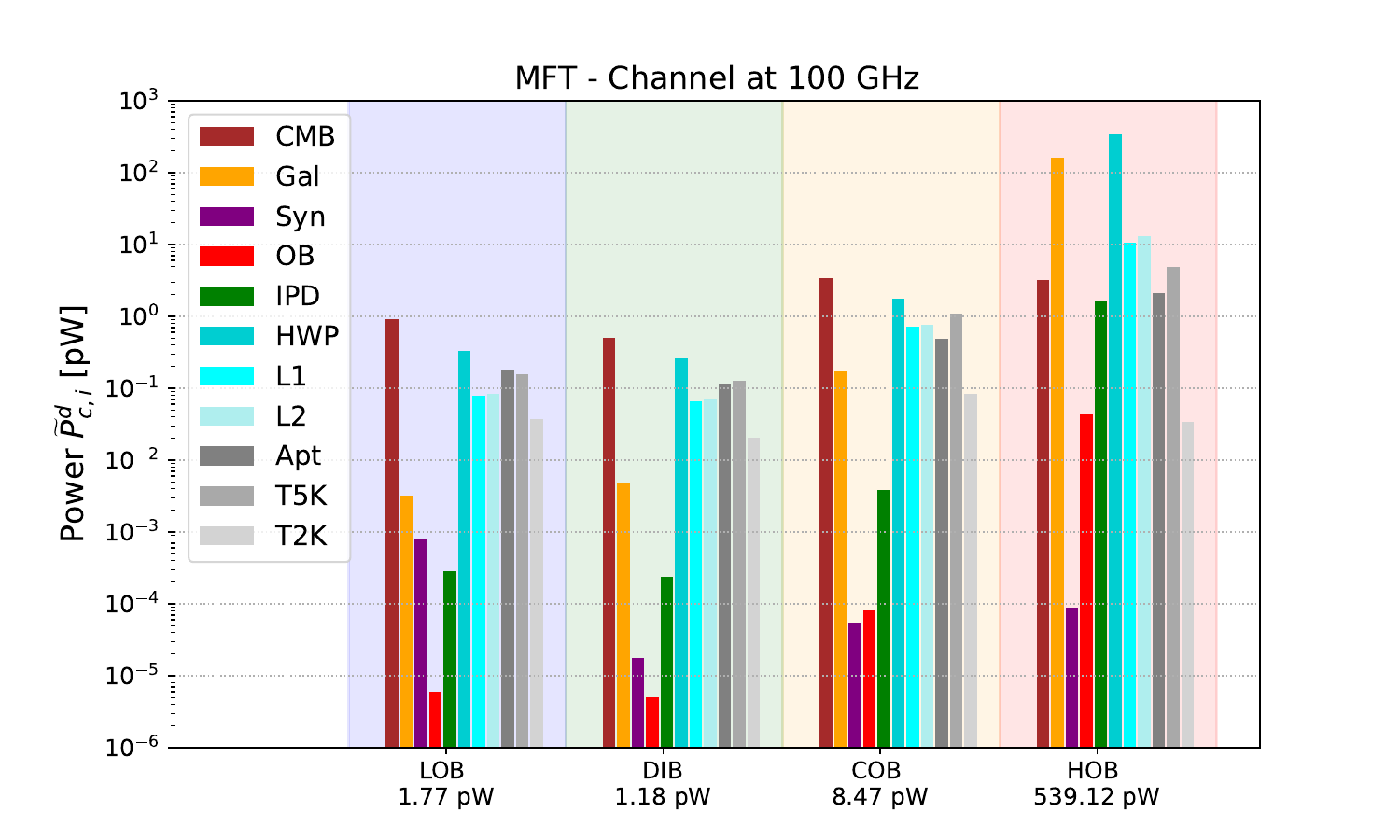}
    \caption{Incident power $\widetilde P^d_{c,i}$ on a MFT detector for the 100\,GHz channel. The x-axis is split in four domains (LOB, DIB, COB and HOB), shown with background colours. We consider five contributions from the sky, CMB, Galactic dust (Gal), Inter-Planetary Dust (IPD), stars (OB), synchrotron (Syn) and five emissions from instrument sub-systems (HWP, L1 and L2, the aperture stop (Apt) and tube at 5\,K and the hood at 2\,K). The total power in each domain is written at the bottom while the repartition between all emissive components are shown with bare-plot.}
    \label{fig:power_detector}
\end{figure}

We have calculated the incident power $\widetilde P^d_{c,i}$ on a detector for the LOB, DIB, COB, and HOB domains, coming from the sky and instrumental components $c$. As instrumental components, we have considered optical elements (HWP, L1 and L2) and the mechanical structure (aperture stop, tube at 5\,K and the hood at 2\,K). The result is shown in figure~\ref{fig:power_detector}, considering the 100\,GHz frequency channel of the MFT. The x-axis is split in four domains shown as background colours. The total power received in each domain is written at the bottom while the repartition between all emissive components are shown with bare-plot. This computation was done for the 10 MFT and HFT frequency channels and a complete table with the details per component can be found in appendix~\ref{app:det_power}.

As a first statement, we note that the total power in the HOB domain is about three orders of magnitude above the total power in the three other domains. This is mainly due to the HWP emission because of its high temperature of 20\,K and to the Galactic dust emission. Therefore, we can already anticipate that we will need to efficiently filter the HOB domain.

% ------------------------------------------------------------
\subsubsection{Requirements from detector NEP performance}
% ------------------------------------------------------------
\label{sss:NEP}

%%% Input: load power from out-of-band from sky and instrument
%%% Requirement Criteria: NEP
%%% Domains d: L C H
%%% Target: $\prod_s A_{i,s}^d$ since we deal with detectors

The increase in incident power on the detectors as a result of the out-of-band contribution impacts the detector NEP. In this study, the channel bandwidth is fixed and we do not aim at optimizing it, we only focus on signal degradation due to out-of-band power. The out-of-band power is fully considered as a contaminant as if it contained no signal of interest. This is why we focus here on the impact of out-of-band on NEP and not on NET.

In this section, we quantify this impact and derive requirements on attenuation factors $A_{c,i}^d$ for $d=L,C,H$. We recall that the requirement is established on the total attenuation factor combining all positions $s$.

The photon NEP is directly derived from the incident power on the detector. The calculation is shown in appendix~\ref{app:NEP}. In order to estimate the total detector NEP, we consider several contributions with the following budget allocation. The total NEP is split in an internal and an external contribution such as
\begin{equation}
    \label{eq:NEP_tot}
    \rm{NEP}_{det} = \sqrt{\rm{NEP}_{int}^2 + \rm{NEP}_{ext}^2}.
\end{equation}
The internal part includes photon NEP, thermal carrier NEP and readout NEP. The external NEP includes noise due to vibrations of the focal plane, thermal fluctuations of the bath temperature, TES instability caused by cosmic ray hits, magnetic flux fluctuations across the TES and electromagnetic interference within the readout system. The procedure for computing the total NEP is described in appendix~\ref{app:NEP} and more details can be found in~\cite{Hasebe_2023}. In table~\ref{tab:NEP} we give the detector NEP obtained for each MFT and HFT frequency channel.

%%% Requirement
To set the requirement for an increase in NEP due to out-of-band power, we base ourselves on its impact on the precision of the $r$ measurement. To do this, we calculated the derivative of the statistical uncertainty $\sigma(r)$ with respect to the detector's NEP for each frequency channel independently. This leads to an acceptable $\Delta \rm{NEP}$ between 0.14 and \SI{2.6}{\atto \watt \per \sqrt \hertz}, as a function of the frequency channel. This estimation relies on a specific component separation method. Also, as this paper aims at setting requirements in the most general manner, and as safely as possible, we have decided to set the requirement to a single value for all bands, the most constraining one, i.e. $\Delta \rm{NEP} \leq\, $\SI{0.14}{\atto \watt \per \sqrt \hertz}.

\begin{figure}[ht!]
    \centering
    \includegraphics[width=.48\linewidth]{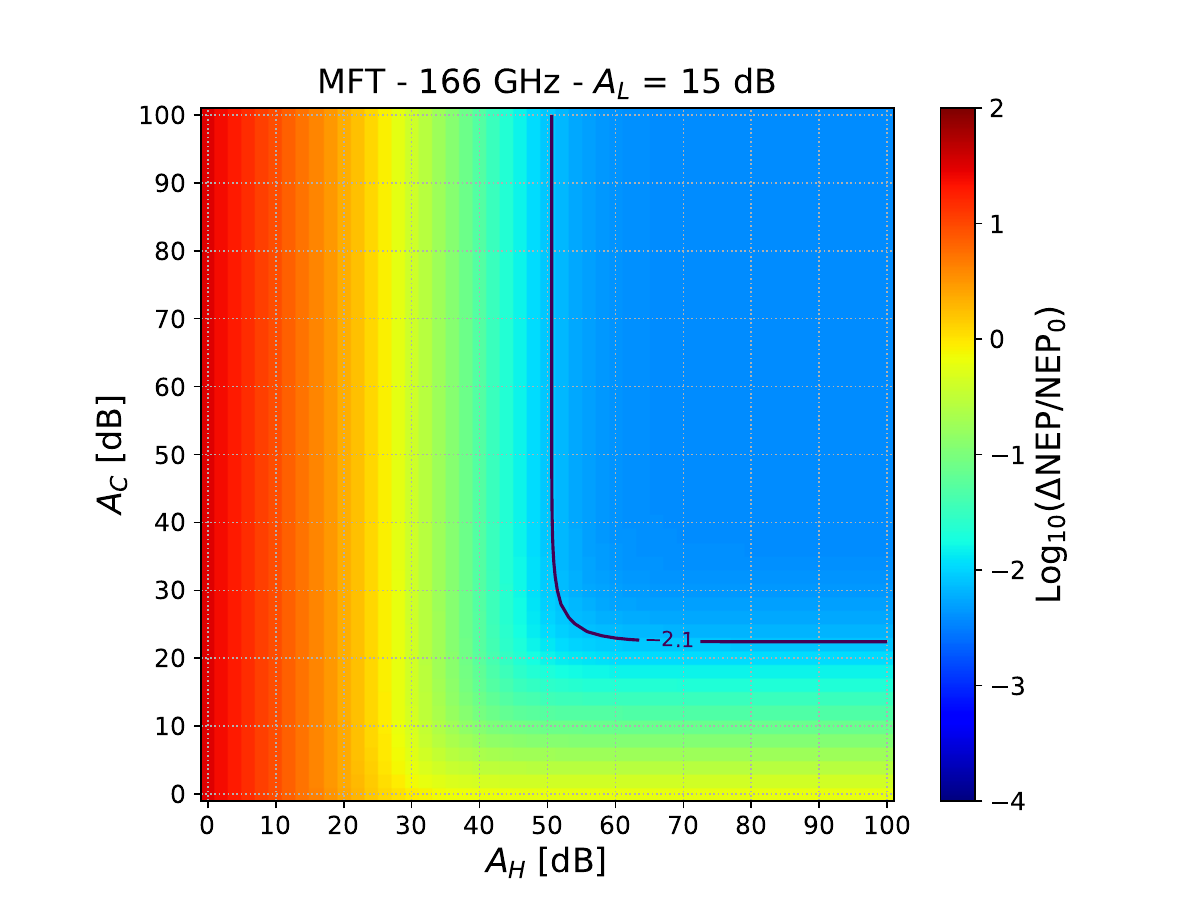}    
    \includegraphics[width=.48\linewidth]{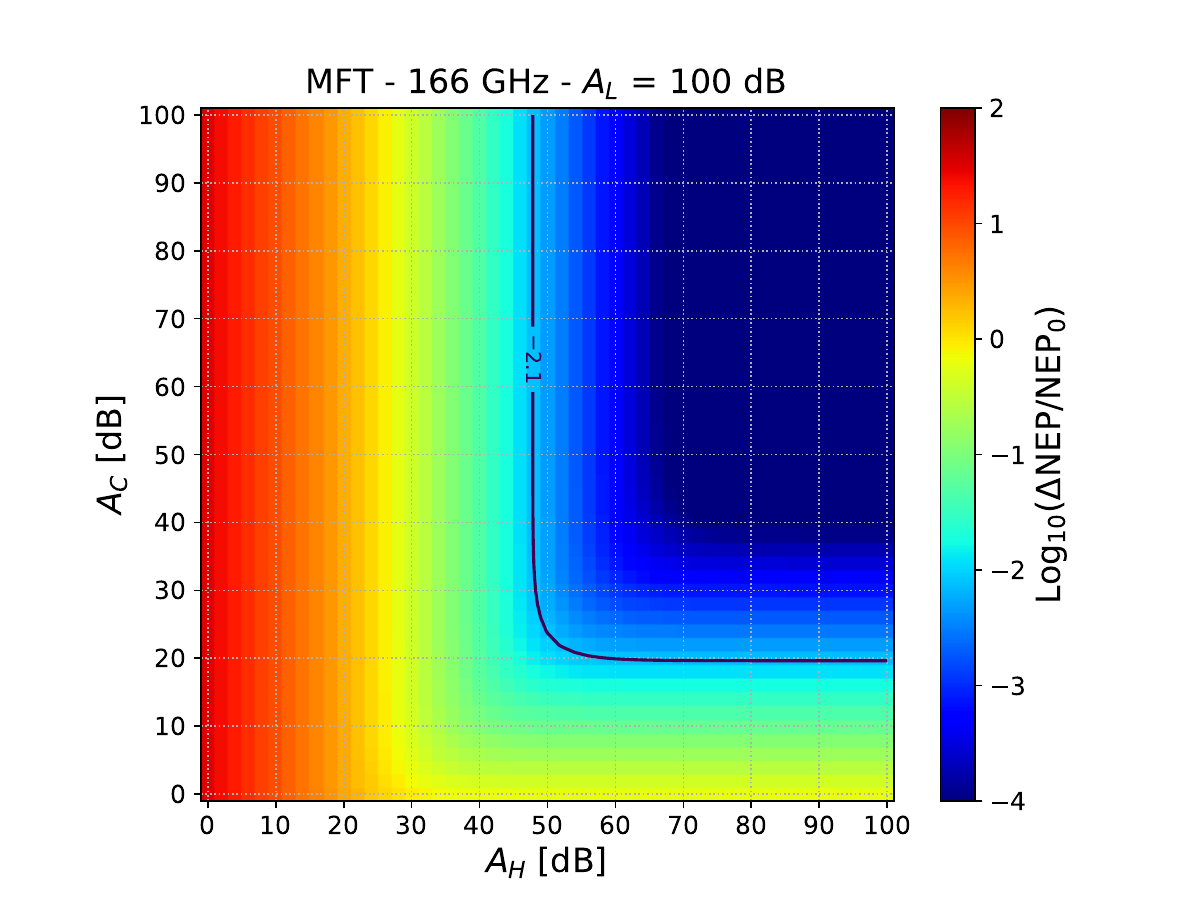} 
    \caption{Relative NEP shift in logarithmic scale, $\log_{10}(\delta_{\rm NEP})$, for a set of attenuation factors in case of the 166\,GHz MFT channel. We have fixed $A_L$ to 15\,dB (left) and 100\,dB (right) and we vary $A_C$ and $A_H$ on a thin grid between 0 and 100\,dB. The requirement on the relative NEP shift is shown as a black line, here -2.1 for this channel.}
    \label{fig:NEP}
\end{figure}

The impact on the NEP is quantified by the relative NEP shift defined as
\begin{equation}
    \label{eq:NEP_shift}
    \delta_{\rm NEP} = \frac{\Delta \rm{NEP}}{\rm{NEP}_0} = \frac{\rm{NEP}_1 -\rm{NEP}_0}{\rm{NEP}_0},
\end{equation}
where $\rm{NEP}_0$ is the reference detector NEP assuming no out-of-band contamination while $\rm{NEP}_1$ is the detector NEP adding the out-of-band contribution. We have evaluated the NEP shift for a set of attenuation factors in the LOB, COB and HOB domains, meaning that the received power by a detector in channel $i$ is
\begin{equation}
    \label{eq:power_attenuation}
    P_{i} = \sum_c \left (\widetilde P_{c, i}^D + A_i^L \widetilde P_{c, i}^L + A_i^C \widetilde P_{c, i}^C + A_i^H \widetilde P_{c, i}^H \right ) \,.
\end{equation}
In figure~\ref{fig:NEP}, we show the NEP shift for a set of attenuation factors in case of the 166\,GHz MFT channel. In this example, we have fixed $A^L_i$ and we varied $A^C_i$ and $A^H_i$ on a thin grid between 0 and 100\,dB. The two figures compare the evolution of the shift for two extreme values of $A^L_i$: 15 and 100\,dB. The requirement on the relative NEP shift is shown as a black line. As we can see, improving a lot $A^L_i$ does not significantly reduce the constraints on $A^C_i$ and $A^H_i$. 

We ran similar calculation for the ten MFT and HFT frequency channels, varying $A^L_i$, $A^C_i$ and $A^H_i$ on a thin grid, allowing us to derive requirements for each frequency channel which are presented in table~\ref{tab:NEP}. As expected, the most stringent requirements are for the HOB domain, reaching between 58 and 82\,dB. 

\begin{table}[ht!]
\centering
\begin{tabular}{|c|c|c|c|c|c||c|c|c|c|c|}
\hline
& \multicolumn{5}{c||}{\textbf{MFT}} & \multicolumn{5}{c|}{\textbf{HFT}} \\
\hline
\textbf{Channel $i$ [GHz]} & 100 & 119 & 140 & 166 & 195 & 195 & 235 & 280 & 337 & 402 \\
\hline
\textbf{$\rm{NEP}_0$ [\SI{}{\atto \watt \per \sqrt \hertz}]}& 10.64 & 12.25 & 12.52 & 12.68 & 12.77 & 17.19 & 15.77 & 15.04 & 14.42 & 13.34 \\
\hline
\textbf{$A^L_i$ [dB]}& 17 & 28 & 20 & 15 & 19 & 23 & 16 & 23 & 22 & 20 \\
\hline
\textbf{$A^C_i$ [dB]}& 34 & 46 & 36 & 30 & 18 & 28 & 42 & 24 & 22 & 52 \\
\hline
\textbf{$A^H_i$ [dB]}& 82 & 66 & 66 & 68 & 76 & 72 & 78 & 80 & 72 & 58 \\
\hline
\end{tabular}
\caption{Requirements on the three attenuation factors $A_i^L, A_i^C, A_i^H$ for the ten MFT and HFT channels, derived from the requirement $\Delta \rm{NEP} \leq\, $\SI{0.14}{\atto \watt \per \sqrt \hertz}. We also give the reference detector NEP for each frequency channel.}
\label{tab:NEP}
\end{table}

% ------------------------------------------------------------
\subsubsection{Requirements from component separation efficiency}
% ------------------------------------------------------------
\label{sss:comp_sep}

%% Input: load power from out-of-band from polarised sky only 
%% Requirement Criteria: r through component separation
%% Domains d: L C H
%% Target: $A_i^d = \prod_s A_{i,s}^d$ since we deal with detectors

In this section we aim at characterizing the impact of  out-of-band power on the component separation between CMB signal and astrophysical foregrounds. \textit{LiteBIRD} will measure the polarized fraction of the signal. Given that the instrumental power contributions are not modulated by the HWP, and that the HWP is the first optical element on the light path, they can not be mixed up with the CMB signal. This is why, in this section, we focus on sky components only.

\begin{figure}
    \centering
    \includegraphics[width=1.\linewidth]{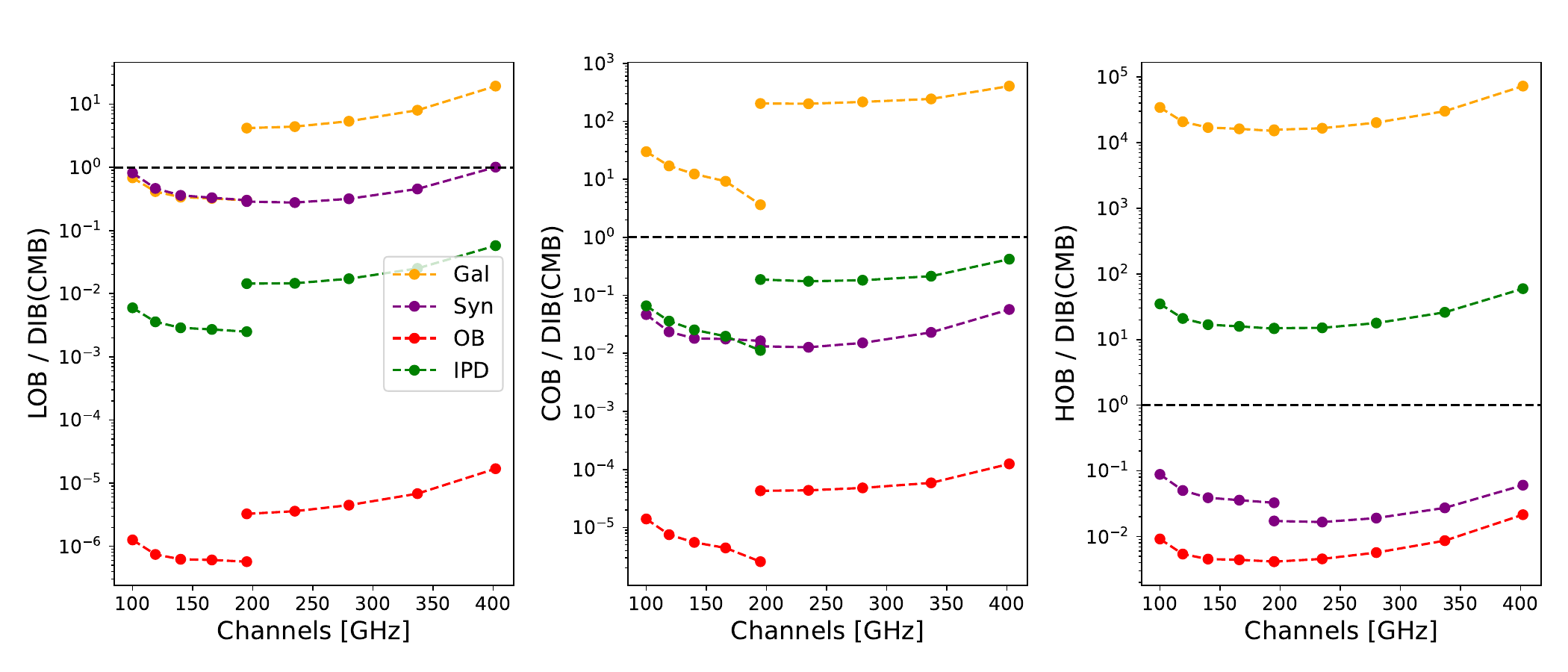}    
    \caption{Ratio of each sky component emitted power, in the LOB (left), COB (middle) and HOB (right) domains over the CMB power received in the detector band (DIB). Ratio are computed considering only the polarized fractions of each component. The black dash line indicates where the ratio is 1. Each colour is associated to a sky component: Galactic dust, Synchrotron, OB stars and IPD. We have split the line between MFT and  HFT channels.}
    \label{fig:ratio_CMB_DIB}
\end{figure}

For each LOB, COB and HOB domain we have computed the ratio between the power emitted by each sky component and the CMB power received in the detector band (DIB). These ratios are computed only considering the polarized fractions of each component, given in section~\ref{ss:skymodel}. The  result is shown in figure~\ref{fig:ratio_CMB_DIB} for each MFT and HFT frequency channel. This clearly shows that the Galactic dust emission, shown in yellow on the figure, is the main issue for component separation. For MFT channels, looking at the ratio LOB over CMB in the DIB domain (left plot), we see that synchrotron emission is as strong as Galactic dust, which is expected at low frequency. This study led us to deeper explore the impact of out-of-band power from synchrotron and Galactic dust on component separation.

In order to do that, we derive requirements based on a complete component separation method as presented in \cite{Stompor:2016hhw} and implemented in the \texttt{xForecast} routine of the \texttt{FGBuster}\footnote{\url{https://fgbuster.github.io/fgbuster/index.html}} library. \texttt{xForecast} requires as inputs the observed (or simulated) frequency maps, an instrumental configuration (i.e. frequency bands and their sensitivities) and a parametric scaling law per sky component, parametrized by what we call spectral parameters. The routine first numerically recovers the spectral parameter values through the maximum likelihood principle and then the component separated maps and the corresponding power spectra. In particular the power spectra for the foreground residuals and noise contributions in the recovered CMB are computed analytically and then used for the cosmological analysis step that, through a Gaussian likelihood, gives estimates of $r$ and its statistical uncertainty $\sigma(r)$. We present in the following the procedure followed to derive the requirements.

As frequency maps to use in the component separation we take the thermal dust emission and the synchrotron emission maps as provided by the models d0 and s0 of the package \texttt{PySM3}~\cite{Thorne:2016ifb}, in accordance to the sky model presented in section~\ref{ss:skymodel}. In the following these maps are called \textit{nominal frequency maps}. These foreground models are considered to be the simplest models currently available in the literature and optimistic with respect to the real foregrounds. Nonetheless they are a suitable choice here, as the goal of this work is not to asses the performances of the component separation technique against complex foregrounds, but rather to evaluate the impact of the presence of out-of-band power on the component separation procedure and hence on the forecasted $r$ value.

We then build the total out-of-band contribution maps associated to each frequency channel $i$ and for each domain $d$ (LOB, COB, HOB). This is done by considering the out-of-band powers presented in section~\ref{sss:power_det}, summed over the three $c$ components (CMB, Galactic dust and synchrotron) and divided by the DIB power. Such ratios are then multiplied by each nominal frequency map. In the following the resulting rescaled maps are called \textit{excess maps}.

At this point a first analysis is performed, considering additional out-of-band power only in one frequency channel $i$ and in one domain $d$ (LOB, COB or HOB). To one nominal frequency map, we add a fraction $x$ of the excess map and we forecast the corresponding $r_{\mathrm{bias}}$. We then repeat the procedure for many values of $x$. We check that for $x=0$ we recover $r_{\mathrm{bias}}=0$, thus the bias on $r$ obtained for a non zero $x$ fraction is only due to the presence of out-of-band power. We look for the largest value of $x$ that gives a bias on $r$ satisfying the requirement value ($6.5 \times 10^{-6}$ in the case of \textit{LiteBIRD}) divided by the number of domains (LOB, COB, HOB) and the number of frequency bands (22 in the case of \textit{LiteBIRD}). This fraction is called $x_{\rm req}$, and we recover one $x_{\rm req}$ per channel. 

This first analysis is a bit too simple as it treats the out-of-band power for each frequency channel and each out-of-band domain independently, we instead want to evaluate the impact of having all of them simultaneously. Therefore, we reapply the \texttt{xForecast} routine on the input frequency maps having an excess in all frequency bands and all domains at the same time. The excess fractions considered are the $x_{\rm req}$ computed in the simple analysis. By construction, this gives a bias on $r$ larger than what is required, so we iteratively rescale all the $x_{\rm req}$ values until the requirement on $r$ is satisfied. These new values lead to the requirements on the attenuations factors $A_i^d$ presented in table~\ref{tab:req_comp_sep}. We are aware that considering all frequency bands and domains affected simultaneously by the out-of-band power is the most pessimistic configuration and we choose it on purpose, to be conservative in setting the requirements.

\begin{table}
    \centering
    \begin{tabular}{|c|c|c|c|c|c||c|c|c|c|c|}
        \hline
        & \multicolumn{5}{c||}{\textbf{MFT}} & \multicolumn{5}{c|}{\textbf{HFT}} \\
        \hline
         \textbf{Channel $i$ [GHz]} & 100 & 119 & 140 & 166 & 195 & 195 & 235 & 280 & 337 & 402 \\
         \hline
         \textbf{$A^L_i$ [dB]} & 40 & 38 & 40 & 42 & 39
                               & 36 & 34 & 27 & 33 & 32 \\
         \hline
         \textbf{$A^C_i$ [dB]} & 44 & 41 & 34 & 38 & 34 & 
                                 47 & 45 & 41 & 44 & 39 \\
         \hline
         \textbf{$A^H_i$ [dB]} & 69 & 66 & 64 & 60 & 69 &
                                 66 & 64 & 60 & 65 & 62 \\
         \hline
    \end{tabular}
    \caption{Requirements found on the attenuation factors from the presence of out-of-band power in the component separation, resulting on a bias on the tensor-to-scalar ratio measurement.}
    \label{tab:req_comp_sep}
\end{table}

%---------------------------------------------------------------------
\subsubsection{Realistic HWP modelling} 
%---------------------------------------------------------------------
\label{sss:HWP_margin}
In this work, no instrumental systematic effect was considered. However we asked ourself if HWP defects could reinforce the out-of-band contamination so that the derived requirements would not be strong enough.  

This is why we performed an additional analysis which allows us to include the non-ideality of the HWP when estimating the impact of out-of-band contamination on the bias on $r$ estimated through component separation. Differently from the one presented in the previous section this analysis has to be based on time-order data (TOD) which corresponds to the signal acquired scanning the sky. The \textit{LiteBIRD} collaboration has built a software, called \texttt{LiteBIRD-sim}\footnote{\url{https://LiteBIRD-sim.readthedocs.io/en/master/index.html}}, to simulate TOD and this is what we have used here. 

\begin{figure}[ht!]
    \centering
    \includegraphics[width=0.7\linewidth]{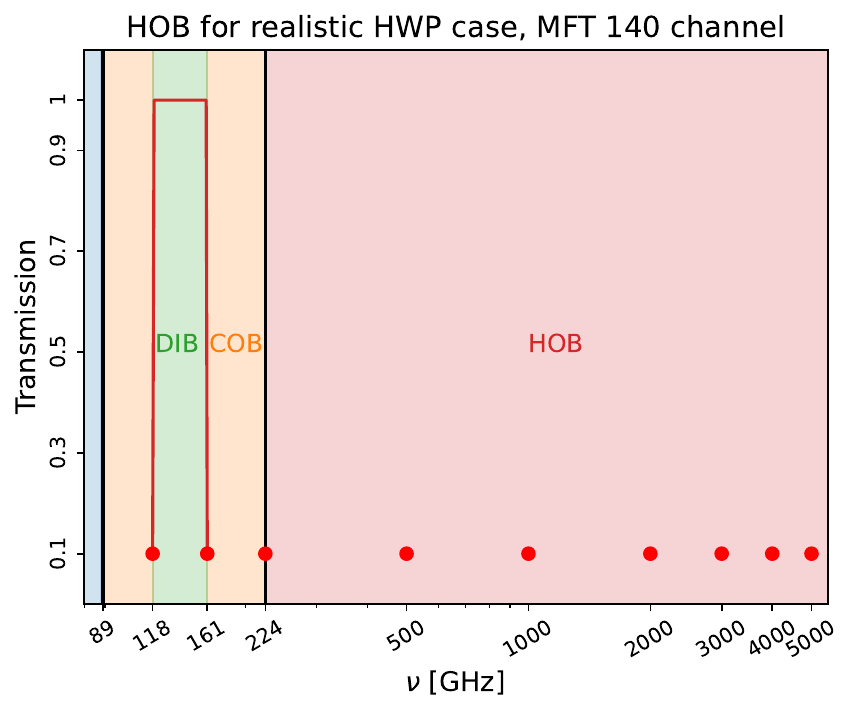}    
    \caption{Example of the HOB case with realistic HWP, for the MFT 140\,GHz channel. The transmission is a simple top-hat in the DIB domain (red line), plus a constant excess (discrete red points) set at 0.1 in this plot. The DIB, COB, LOB and HOB regions are represented by the shaded blue, orange, green, and red areas.}
    \label{fig:hwp_margin_domain}
\end{figure}

The out-of-band effect is modelled as a constant excess in the bandpass profiles out of the DIB frequency range. We consider two cases:
\begin{itemize}
    \item the equivalent of the COB case, in which we assume a top-hat bandpass profile for the DIB frequency domain and a constant non-null bandpass excess in the other frequencies of the telescope (which corresponds to the HWP nominal frequency range);
    \item the equivalent of the HOB case, shown in figure~\ref{fig:hwp_margin_domain}. We use a top-hat bandpass profile in the DIB plus a constant excess at few discrete points placed at the edges of the DIB, at the higher edge of the COB, and at very high frequencies at 500, 1000, 2000, 3000, 4000 and 5000\,GHz. 
\end{itemize}

We parametrize the HWP through the Jones matrix $J_{\rm HWP}$, describing how the incoming electromagnetic waves are affected by the optical element \cite{ODea:2006tvb}:
\begin{equation} \label{eq:realistic}
\begin{pmatrix}
E_\mathrm{x, out} \\ E_\mathrm{y, out}
\end{pmatrix} = 
J_{\rm HWP} \begin{pmatrix}
E_\mathrm{x, in} \\ E_\mathrm{y, in}
\end{pmatrix} = \begin{pmatrix}
    1+h_{1} & \zeta_{1} \\
    \zeta_{2} & -(1+h_{2}) e^{i \beta} \\ 
\end{pmatrix} \begin{pmatrix}
E_\mathrm{x, in} \\ E_\mathrm{y, in}
\end{pmatrix}.
\end{equation}
The meaning of the non-ideal parameters in $J_{\rm HWP}$ is as follows: 
\begin{itemize}
        \item[$\bullet$]  $h_{1}$ and $h_{2}$:  loss parameters describing the deviation from the unitary transmission of light components $E_{x}$, $E_{y}$;
        \item[$\bullet$] $\beta = \phi - \pi $: where $\phi$ is the phase shift between the two directions. It accounts for variations of the phase difference between $E_{x}$ and $E_{y}$ with respect to the nominal value of $\pi$ for an ideal HWP;
        \item[$\bullet$] $\zeta_{1,2}$:  amplitudes of the off-diagonal terms, coupling $E_{x}$ and $E_{y}$. In practice, if the incoming wave is fully polarized along $x$ ($y$), a spurious $y$ ($x$) component would show up in the outgoing wave. 
\end{itemize}

In the ideal case, all these parameters are null. In general, the non-ideal HWP parameters are frequency dependent. We adopt a model\footnote{See Figures 5 and 6 of~\cite{Giardiello:2021uxq} for the simulated $h$ and $\beta$ profiles of the MFT channels.} of the  metal-mesh HWP derived for the MHFT channels~\citep{LiteBIRD:2020zfx}. In a simulated Mesh-HWP the cross polarization parameters $\zeta$ are exactly null because of the supposed exact symmetry of the system. Because of this, there are no simulated frequency profiles for $\zeta_1, \zeta_2$ and so, for simplicity, we assumed them to be constant and equal to $10^{-2}$. In the case of LFT, since we have no simulations for the LFT HWP non-ideal parameters, we just consider the case of an ideal HWP. This is justified by the fact that we do not find a relevant difference in the bias on $r$ in the MHFT channels due to assuming a realistic or an ideal HWP.

In this part of the analysis, we compute a TOD and perform map-making in the presence of different level of constant COB/HOB excess, both assuming an ideal and a realistic HWP, for all the frequency channels. The level of COB excess spans from 0.1 to $10^{-5}$ for LFT and MFT, and from 0.1 to $10^{-7}$ for HFT. For HOB, we explore excesses from 0.1 to $10^{-9}$ for all the channels. The details of this procedure are described in appendix~\ref{app:HWP_margin}. 

For each excess, we build a residual map which is then summed to the nominal frequency maps as defined in section~\ref{sss:comp_sep} and fed in the \texttt{xForecast} component separation routine. We thus recover the bias on the tensor-to-scalar ratio from these maps and we compare it with the values recovered from the analysis in section~\ref{sss:comp_sep} in case of input maps with the same excess in the two approaches. We find that the additional bias due to the out-of-band power for the realistic HWP is negligible with respect to the values coming from the analysis in~\ref{sss:comp_sep}. Therefore, we conclude that the HWP non-ideality is negligible at this stage of the requirement forecasting, given the current HWP configuration. This may change for HWP simulations where the deviations from ideality are more pronounced.

%%%%%%%%%%%%%%%%%%%%%%%%%%%%%%%%%%%%%%%%%%%%%%%%%%%%%%
\subsection{Requirements from thermal cooling budget - Static analysis} 
%%%%%%%%%%%%%%%%%%%%%%%%%%%%%%%%%%%%%%%%%%%%%%%%%%%%%%
\label{ss:req_from_thermal}

In this section, we quantify the impact of out-of-band power on the thermal balance of the instrument. Additional heat load must comply with the cooling instrument power capability both on the focal plane, $P_{\rm det}^{\rm cool}$, and on optical elements (L1, L2, and the HWP), $P_{\rm inst}^{\rm cool}$. These constraints allow us to derive requirements on the attenuation factors for the LOB and HOB domains.

%---------------------------------------------------------------------
\subsubsection{Radiative heat load computation} 
%---------------------------------------------------------------------
\label{sss:heat_load}

\begin{figure}
    \centering
    \includegraphics[width=1\linewidth]{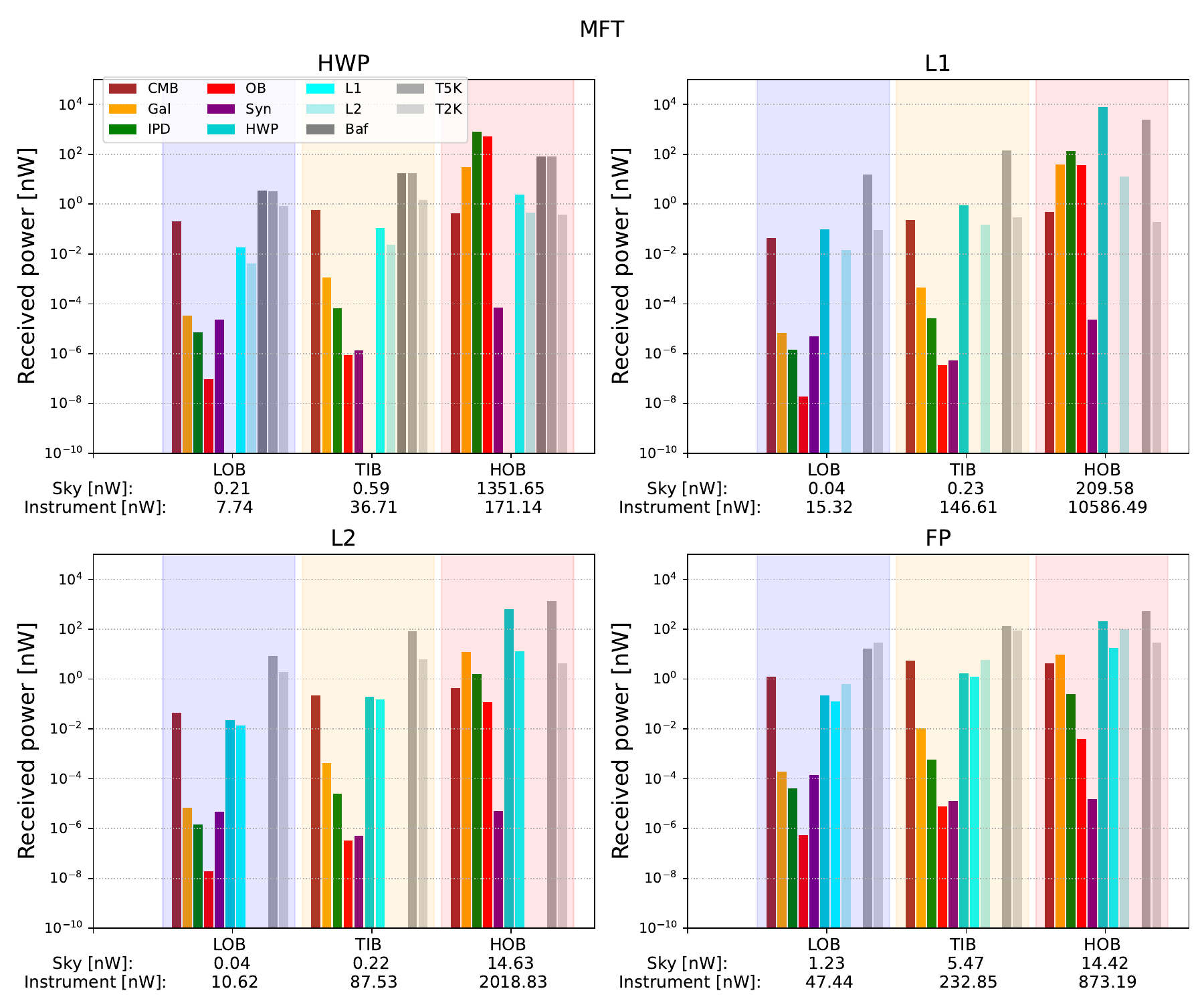}    
    \caption{Heat load on the HWP, the two lenses L1 and L2 and the focal plane FP for the MFT separated in three frequency domains: LOB (blue background color), TIB (yellow) and HOB (red). Each colour bar is associated to a given contribution either from the sky or from the instrument itself. The total power from the sky and from the instrument in each frequency domain is summarized below each plot in nW.}
    \label{fig:radiative_heat_load_MFT}
\end{figure}

\begin{figure}
    \centering
    \includegraphics[width=1\linewidth]{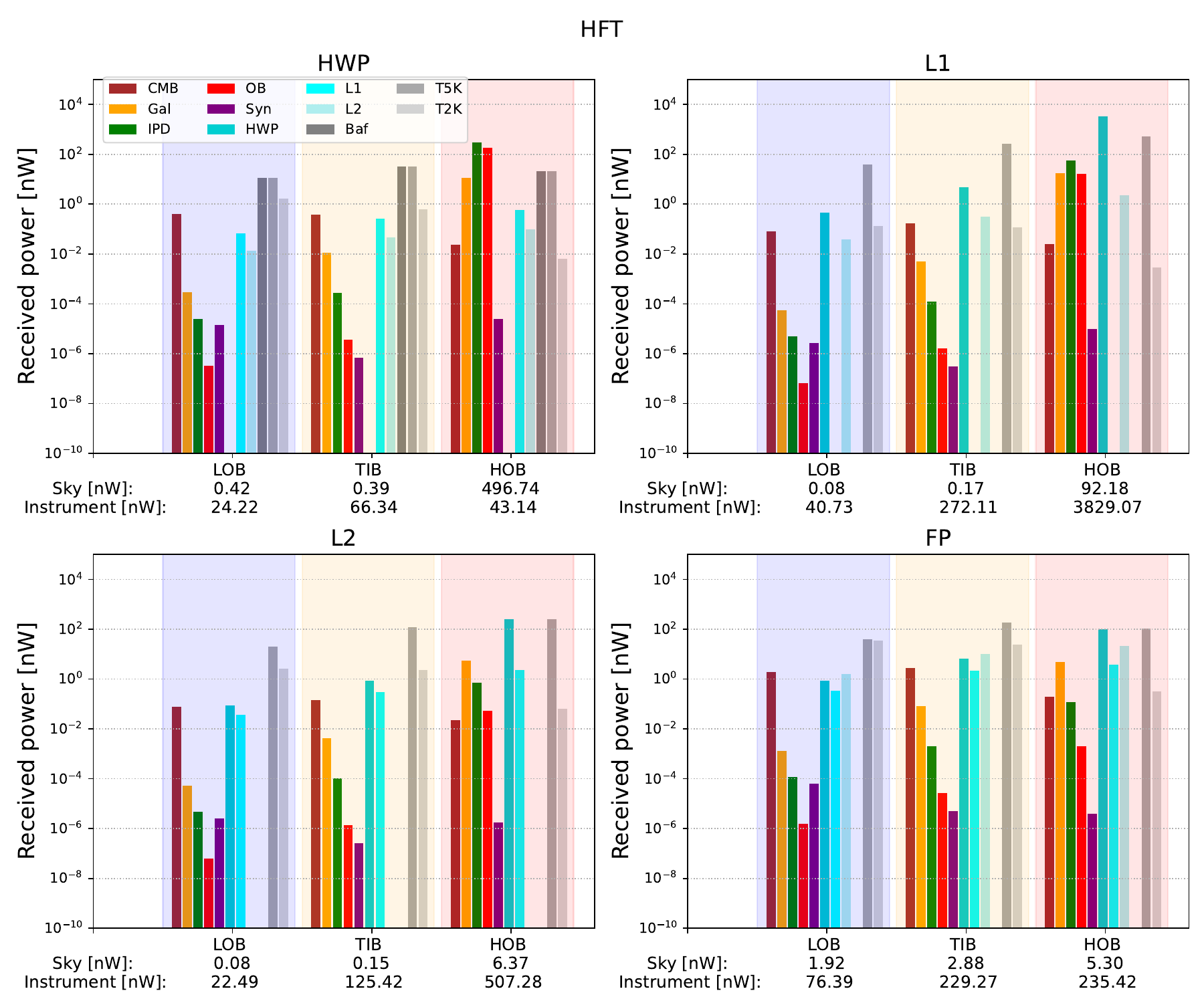}
    \caption{Similar to figure~\ref{fig:radiative_heat_load_MFT} but for the HFT.}
    \label{fig:radiative_heat_load_HFT}
\end{figure}

We have evaluated the radiative heat load on the HWP, the two lenses L1 and L2, and the focal plane (FP), both from the sky components and from the instrument itself, considering the LOB, TIB and HOB frequency domains defined in section~\ref{ss:band_def}. The total heat load on each element is calculated as described in section~\ref{sss:radiative_load}, and the result is shown in figure~\ref{fig:radiative_heat_load_MFT} for MFT and figure~\ref{fig:radiative_heat_load_HFT} for HFT. We neglect the thermal load coming from the baffle, except onto the HWP which is adjacent to it.

% Puissance totale dominée par le HOB et par l'instrument 
The total heat load is clearly dominated by the HOB domain. As we can see on figures~\ref{fig:radiative_heat_load_MFT} and~\ref{fig:radiative_heat_load_HFT}, HOB heat load is about two or three orders of magnitude higher than the LOB or TIB contributions. Moreover, we note that the sky contribution on total heat load is subdominant compared to the instrument contribution, except on the HWP in the HOB domain.

%---------------------------------------------------------------------
\subsubsection{Heat load on the focal plane} 
%---------------------------------------------------------------------
\label{sss:FP_heat_load}

\begin{table}[ht!]
    \centering
    \begin{tabular}{|c|c|c||c|c|}
        \hline
        & \multicolumn{2}{c||}{\textbf{MFT}} & \multicolumn{2}{c|}{\textbf{HFT}} \\
        \hline
         &  \textbf{$A^L$ [dB]} & \textbf{$A^H$ [dB]} &  \textbf{$A^L$ [dB]} & \textbf{$A^H$ [dB]}  \\
         \hline
         $\Delta P_{sky}$ & / & 8 & / & 4 \\
         \hline
         $\Delta P_{HWP}$ & / & 19 & / & 16 \\
         \hline
         $\Delta P_{L1}$ & / & 9 & / & 2  \\
         \hline
         $\Delta P_{L2}$ & / & 16 & / & 10  \\
         \hline
         $\Delta P_{T5K}$ & 9 & 23 & 12 & 17  \\
         \hline
         $\Delta P_{T2K}$ & 11 & 11 & 12 & /  \\
         \hline
    \end{tabular}
    \caption{Requirements on $A^L$ and $A^H$ from thermal heat load onto the FP for MFT and HFT. The sky components are grouped together while we detail the instrument contributions (HWP, L1, L2, T5K and T2K). The backslash means that the component does not need to be filtered, see the text for more details.}
    \label{tab:FP_heat_load}
\end{table}

The detector operating temperature is~\SI{100}{\milli \kelvin} and the maximum heat load on each FP (LFT, MFT and HFT) is~\SI{2.2}{\micro \watt} including parasitic heat load, harness, radiative loading, etc. Since we deal with the FP, we cannot constrain the attenuation factor for position $s_0$ that would correspond to the on-chip filter. Therefore, we expect to put requirements on the attenuation factors $A_s^d$ for $d=L, H$ and $s = s_1, s_2, s_3, s_4$. We note that there is no channel dependency $i$ as the radiative heat load computation is not specific to a channel.

The radiative heat load on the FP is shown in figures~\ref{fig:radiative_heat_load_MFT} and~\ref{fig:radiative_heat_load_HFT}, bottom right plots. Regarding the total power (sky + instrument) from TIB we have about~\SI{0.24}{\micro \watt} for MFT and same for HFT. The TIB domain can, of course, not be attenuated and represents about 11\% of the total budget. 

The requirement we consider is the following: the heat load contribution from LOB and HOB domains from each sky or instrumental component should not be more than 1\,\% of the total heat load in the TIB domain. We treat all the sky components together, however, we detail a requirement for each instrumental contribution. This allows us to impose constraints on the optimum position of the filters $s_1, s_2, s_3, $ or $s_4$. Requirements on $A^L$ and $A^H$ are presented in table~\ref{tab:FP_heat_load} for MFT and HFT. The backslash means that this component does not need to be filtered as it is already below the requirement. 

We see that the requirements we obtained are not very stringent. For the LOB domain, the only relevant contributions are from the tubes at 5\,K and 2\,K. For the HOB domain, the most relevant contributions are the HWP, the tube 5K and the lens L2. This favours to place a filter at position $s_1$, between lens L2 and the focal plane rather than earlier on the optical path ($s_2, s_3, s_4$). 

One additional remark is that for this analysis, we have considered that all the power received by the focal plane is absorbed. In reality, the surface of the focal plane will be largely reflective, making this analysis pessimistic. For example, we have assumed that 30\% of the incident power is reflected, which relaxes the requirements presented in Table~\ref{tab:FP_heat_load} by 1 to 2 dB each. As the design of the focal plane has not yet been finalized, it is difficult to give a value for reflectivity. For this reason, we prefer to consider that all the received power is absorbed, bearing in mind that this assumption may be refined in the future.

%---------------------------------------------------------------------
\subsubsection{Heat load on the HWP} 
%---------------------------------------------------------------------
\label{sss:HWP_heat_load}

Radiative heat load on the HWP is shown in figures~\ref{fig:radiative_heat_load_MFT} and~\ref{fig:radiative_heat_load_HFT}, top left plots. The heat load is clearly dominated by the HOB domain. Due to the strong absorption features of polypropylene in the thermal infra-red, and its high transparency at long wavelengths, the most relevant heat source for the HWP are IPD and OB star emissions. Instrumental emissions in the HOB domain are about one order of magnitude below.

These estimates should be compared to the total heat load from Eddy currents, which is expected to be of the order of few mW. Indeed, according to current requirements, the total thermal load of the rotator must be less than 7\,mW, and only a portion of it will be dissipated in the rotor. However, this number remains significantly higher than the estimates shown in figures~\ref{fig:radiative_heat_load_MFT} and~\ref{fig:radiative_heat_load_HFT}. This allows us to conclude that an infra-red filter at position $s_4$ in front of the HWP is not necessary from the point of view of total heat load.

%---------------------------------------------------------------------
\subsubsection{Heat load on the lenses} 
%---------------------------------------------------------------------
\label{sss:lenses_heat_load}

% Lentilles
Lenses are cooled at 4.8\,K. Taking into account 33\,\% margin for the mechanical cooler, the available cooling power at the 4.8-K stage is $P_{\rm L1, L2}^{\rm cool} \sim \SI{30}{\milli \watt}$~[PTEP]. This is much higher than the power absorbed by the lenses which is about~\SI{11}{\micro \watt} for L1 and~\SI{2.1}{\micro \watt} for L2 for MFT and~\SI{4.2}{\micro \watt} for L1 and~\SI{0.66}{\micro \watt} for L2 for HFT. Thus, the total heat load falling on the lenses is not an issue and does not imply constraints on the filtering scheme.

%%%%%%%%%%%%%%%%%%%%%%%%%%%%%%%%%%%%%%%%%%%%%%%%%%%%%%
\subsection{Requirements from a dynamic analysis} 
%%%%%%%%%%%%%%%%%%%%%%%%%%%%%%%%%%%%%%%%%%%%%%%%%%%%%%
\label{ss:req_from_dynamic}

This section addresses the impact of thermal variations $\delta P$ along the scanning both from sky intensity variations and from instrument thermal fluctuations. 

As mentioned in section~\ref{sss:HWP_margin}, we can simulate TOD using the \texttt{LiteBIRD-sim} software. This is what we have used to get the temporal variations from the sky. The scanning period will be approximately 20~minutes during which the satellite will typically pass by the Galactic plane twice~[PTEP]. 

Regarding the instrument's thermal fluctuations, we rely on internal studies and on the thermal characteristics of certain sub-systems~\cite{Sato2021Cryo, Columbro:2020mci}.

We have considered three possible issues that may impact the nominal operational behaviour of the instrument. The first one is a change of the operating detector temperature due to an increase of the focal plane equilibrium temperature. The second one is a variation of the detector NEP due to radiative heat load fluctuations. Finally, the third one is an increase of the mean temperature of the optical elements: the HWP and the two lenses. Those three aspects are studied in the next sections.

% ------------------------------------------------------------
\subsubsection{Requirements from detector NEP stability}
% ------------------------------------------------------------
\label{sss:dynamic_NEP}

%%% Input: fluctuations of load power from out-of-band from sky and instrument
%%% Requirement Criteria: NEP
%%% Domains d: L C H
%%% Target: $\prod_s A_{i,s}^d$ since we deal with detectors

In section~\ref{sss:NEP} we derived requirements from NEP increase due to out-of-band thermal heat load. In this section we address the possibility of additional NEP variations due to thermal fluctuations along the scan from the sky and the instrument.

Thanks to the rotating HWP, the astrophysical signal in polarization is modulated at four times the HWP rotation speed. Therefore, the thermal variations we need to worry about are only the ones included in the signal modulated bandpass $\Delta\nu = [2 - 5]\,\si{\hertz}$. 
Moreover in this section we only address the sky intensity fluctuations and not the instrumental fluctuations because the instrumental ones are not modulated by the HWP. 

%%% SKY
We simulated a 24~hours long TOD for a single detector with a realistic scanning strategy. The map that is scanned is a pure intensity map, including all the sky components described in section~\ref{ss:skymodel}. As we are interested in the detector NEP, the map $I$ has to be smoothed at the detector resolution. We chose to consider the resolution of the highest frequency channel at 402~GHz because it is the channel we most worry about for sky fluctuations as it is the one where the Galactic dust emission is the strongest. Therefore, before computing TOD, the map $I$ is smoothed by a Gaussian with $\rm{FWHM} = 17.9$\,arcmin (see table~\ref{tab:channels}). 

We compute the power spectrum of the TOD and we deduce the additional power in the signal modulated bandpass $\Delta\nu $ due to sky fluctuations. We then add this contribution to the optical power received by the detector which was already computed and we proceed similarly to section~\ref{sss:NEP} in order to set requirements on $A_L$, $A_C$ and $A_H$. As described in annex~\ref{app:NEP}, thermal fluctuations are one source of external NEP among the ones we consider and the budget attributed to each external $\rm{NEP}^2$ is 4.9\,\% of the total detector $\rm{NEP}^2$. 

This leads to the following requirement: $A^C$ and $A^H$ should be at minimum 35\,dB and 15\,dB respectively. This conclusion was established for a low value of $A^L=$14\,dB so requirements would be even less stringent for a stronger $A^L$ attenuation factor.

To be accurate, those requirements apply on the product of the attenuation factors over the filter positions $s_0$ to $s_4$ and for channel $i=402$\,GHz in HFT only. However, we take the same requirement for all channels $i$ (MFT and HFT). This is the most conservative approach as sky fluctuations will be less important for lower frequency channels.

% ------------------------------------------------------------
\subsubsection{Requirements from detector operating temperature stability}
% ------------------------------------------------------------
\label{sss:dynamic_Tdet}

%%% Fluctuations of load power from out-of-band from sky and instrument
%%% Requirement Criteria: FP temperature stability
%%% Constrained domains: L et H
%%% Target: $A_{s1}^d \times A_{s2}^d \times A_{s3}^d \times A_{s4}^d$ since we deal with the whole FP

In this section we set requirements on the out-of-band attenuation factors $A^L$ and $A^H$, from the current constraints on the temperature stability of the FP set to 100\,mK. For that purpose, we consider the LOB, TIB and HOB power falling on the whole FP, already computed in section~\ref{sss:heat_load}. Similarly to section~\ref{sss:FP_heat_load}, since we deal with the whole FP, we set requirements on the product $\prod_{s} A_s^d$ for $d=L, H$ and $s = s_1, s_2, s_3, s_4$ and no channel $i$ dependency.

% Tolerance en W pour les fluctuations sur le PF
From detector gain requirements, we have a tolerance of about \SI{10}{\micro\kelvin} fluctuations on the focal plane at \SI{100}{\milli\kelvin}. This requirement includes all possible sources of fluctuations so we decide to attribute 10\% of it for thermal radiative load variations. In terms of equivalent power, according to a preliminary simplified thermal instrument model developed internally in the \textit{LiteBIRD} collaboration, \SI{1}{\nano\watt} on the focal plane is about \SI{10}{\micro\kelvin} fluctuation. This result may evolve with the level of complexity included in the thermal wafer modelling which has not been performed yet.

%%% SKY
In order to estimate the fluctuations of the mean focal plane temperature due to sky intensity variations we performed the following analysis. We simulated TOD from a pure intensity sky map (no polarized components), smoothed with a Gaussian beam with FWHM equal to $28 \deg$ which is the FP field-of-view. We compute the power spectrum of the TOD which allows us to get the fraction of power due to sky fluctuation intensity which enters in the signal modulated bandpass $\Delta \nu$. From figures~\ref{fig:radiative_heat_load_MFT} and~\ref{fig:radiative_heat_load_HFT}, we already computed the power received by the focal plane from sky in LOB, TIB and HOB. This allows us to know how much the fraction due to sky fluctuation represents in Watt: for MFT(HFT), we obtain about $7.3\times 10^{-3}(1.1\times 10^{-2})$\,nW for LOB,  $3.2\times 10^{-2}(1.7\times 10^{-2})$\,nW for TIB and $8.5\times 10^{-2} (3.1\times 10^{-2})$\,nW for HOB. These contributions are about one order of magnitude below our requirement which is \SI{0.1}{\nano\watt}. Therefore, sky fluctuations are not a concern for detector operating temperature stability.

%%% INSTRUMENT
Concerning the instrument thermal fluctuations, as we look at their impact on the FP temperature, we mainly care about the 2\,K and 5\,K tubes contributions as they are the dominant ones (see figures~\ref{fig:radiative_heat_load_MFT} and~\ref{fig:radiative_heat_load_HFT}). Following the analysis presented in~\cite{Sato2021Cryo} for a similar cooling system, we expect fluctuations of the order of $2\times 10^{-3}\si{\kelvin\per\sqrt\hertz}$, so typically few mK in the signal modulated bandpass $\Delta\nu$. To get the corresponding impact on the FP, we compute the received power following section~\ref{sss:heat_load} for a 3\,mK increase of the 2\,K tube and 5\,K tube temperatures. We obtain that the power increase on the FP coming from the 2\,K and 5\,K tubes is between 0.01 and \SI{0.5}{\nano\watt}, excepted for the MFT in the HOB domain where it reaches about \SI{1.4}{\nano\watt}. However, even in that case, a 20\,dB attenuation factor would be enough. Therefore we conclude that instrumental fluctuations are not critical and low attenuation factors in LOB and HOB domains would be sufficient to satisfy the thermal stability of the FP.

% ------------------------------------------------------------
\subsubsection{Requirements from optical elements thermal stability}
% ------------------------------------------------------------
\label{sss:dynamic_Tinst}

%%% Input: fluctuations of load power from out-of-band from sky
%%% Requirement Criteria: optical element thermal stability
%%% Constrained domains: L et H
%%% Target: $A_{i,s2}^d$ or $A_{i,s3}^d$ since we deal with optical elements

In this section we want to quantify the impact of sky fluctuations on the thermal stability of optical element such as the HWP and the two lenses. We study carefully the impact on the HWP.  Then, we generalize the result to the two lenses. 

Concerning the HWP, the equilibrium temperature is mainly determined by the presence of Eddy currents in the material. Knowing the mechanical and thermal properties of the HWP, we can solve the heat equation and get the temperature of the HWP as a function of time. By doing that we have fit the time constant of the HWP which is about 30~hours. Therefore we can predict that the HWP should behave as a low-pass filter, and that sky fluctuations should be dumped by a large factor.

In order to verify this assumption, we produced TOD from each sky component map, smoothed at the field-of-view resolution which is 28~degrees. Then, we add the TOD to the power received by the HWP and we recompute the evolution of its temperature as a function of time. Since this is computer-time consuming we run this test only for the HOB domain with the Galactic dust emission which is the main cause of thermal fluctuations from sky. This leads to an increase of the HWP equilibrium temperature of the order of $\SI{7}{\micro\kelvin}$ with respect to the case with no additional TOD. This is negligible compared to the mean temperature which is about $\SI{20}{\kelvin}$.

% L1 and L2
Similarly to the HWP, lenses L1 and L2 have a thermal inertia of the order of 10~hours. So for the same reason, sky fluctuations will not translate in time variation at the lens temperature which is 4.8\,K. 

%%%%%%%%%%%%%%%%%%%%%%%%%%%%%%%%%%%%%%%%%%%%%
\subsection{Combination of all requirements}
\label{ss:req_summary}
%%%%%%%%%%%%%%%%%%%%%%%%%%%%%%%%%%%%%%%%%%%%%

We have derived requirements along section~\ref{sec:LiteBIRD} through various analysis. Table~\ref{tab:summary} summarizes the constraints obtained on the attenuation factors $A^L$ (blue), $A^C$ (orange) and $A^H$ (red). Each line refers to a distinct analysis presented in a dedicated section. Let us review the main conclusions derived along the section.

In section~\ref{ss:req_from_det} we derived requirements studying the impact of out-of-band power on the detection chain from a static point of view. We studied both the impact on the detector NEP (section~\ref{sss:NEP}) and on the component separation efficiency (section~\ref{sss:comp_sep}). For both cases, we derived requirements on the three attenuation factors per frequency channel for MFT and HFT. Also, those requirements applied to the total attenuation factor: $A_i^d = \prod_s A_{i,s}^d$ for $d=L, C, H$, which is indicated in table~\ref{tab:summary} (column titled Position $s$). The first thing we can notice is that these two analysis are the most constraining. Also, the most stringent requirements are on the high out-of-band ($A^H$ factor). 

In section~\ref{ss:req_from_thermal}, we studied the impact of out-of-band power on the thermal balance of the instrument. This allowed us to derive requirements on $A^L$ and $A^H$ but not on $A^C$. We first considered the additional heat load on the focal plane (section~\ref{sss:FP_heat_load}). As we can see in table~\ref{tab:summary}, depending on the emitting component (sky or instrument) the combination of attenuation factors we constrain is different. For example, for the sky emission, we constrain the product from position $s_1$ to $s_4$ while for L1 emission, we constrain the product from $s_1$ to $s_2$. Moreover, as we consider the full focal plane, on-chip filter (position $s_0$) is never constrained. As we can see, the requirements we obtained are quite low, reaching at maximum 23~dB for $A^H$. Also, one interesting conclusion of this section was that position $s_1$, between lens L2 and the focal plane, is the most critical because the majority of the heat load in the HOB domain comes from the HWP, the tube 5K and the lens L2. In section~\ref{sss:HWP_heat_load}, we computed the heat load on the HWP and we concluded that it was negligible compared to the power produced by Eddy currents. The heat load on the lenses was also computed in section~\ref{sss:lenses_heat_load}, and we concluded that the received power is negligible compare to the available cooling power.

Finally, in section~\ref{ss:req_from_dynamic} we studied the impact of thermal variations. This is the only section were we consider dynamic variations instead of static additional power. The study of the detector NEP stability (section~\ref{sss:dynamic_NEP}) gave the following requirement: considering $A_L = 14$\,dB, we should at least have $A_C = 35$\,dB and $A_H= 15$\,dB, which is not very stringent. This was established for HFT channel at 402\,GHz but we decided to keep the same requirement for others channels in a conservative approach. We also studied the stability of the detector operating temperature (section~\ref{sss:dynamic_Tdet}) and the thermal stability of the optical elements (section~\ref{sss:dynamic_Tinst}) but we concluded that this was not a constrain for out-of-band rejection levels. In summary, dynamic analysis yields very loose constraints, while static ones are tight. This asymmetry was expected, knowing the long thermal time constants (about 30\,h for HWP and 10\,h for lenses), the fast scan modulation effectively averaging sky variability, and the HWP modulation filtering low-frequency drifts.

\begin{table}
\center
\scriptsize
\begin{tabular}{
      | l | l |
      >{\centering}m{.1cm} |
      >{\centering}m{.1cm} |
      >{\centering}m{.1cm} |
      >{\centering}m{.1cm} |
      >{\centering}m{.1cm} |
      >{\centering}m{.5cm} |
      >{\centering}m{.5cm} |
      >{\centering}m{.5cm} |
      >{\centering}m{.5cm} |
      >{\centering\arraybackslash}m{.5cm} |
      >{\centering}m{.5cm} |
      >{\centering}m{.5cm} |
      >{\centering}m{.5cm} |
      >{\centering}m{.5cm} |
      >{\centering\arraybackslash}m{.5cm} |
      }

\hline
\multicolumn{2}{|c|}{} & \multicolumn{5}{c|}{Position $s$} &  \multicolumn{5}{c|}{MFT $\nu$ [GHz]} &  \multicolumn{5}{c|}{ HFT $\nu$ [GHz]}   \\
\hline
\centering Section & \centering Case & $s_0$ &  $s_1$ &  $s_2$ &  $s_3$ &  $s_4$ &  100 & 119 & 140 & 166 & 195 & 195 & 235 & 280 & 337 & 402 \\

%%%%%%%%% LOB
\hline
\hline
\rowcolor{blue!18}
\multicolumn{17}{|c|}{ $A^{L}$ [dB]} \\
\hline
%-----STATIC
% Detector chain
\rowcolor{blue!10}
\ref{sss:NEP}& $\Delta P \rightarrow NEP$ & \multicolumn{5}{c|}{\cellcolor{gray}} & 17 & 28 & 20 & 15 & 19 & 23 & 16 & 23 & 22 & 20 \\
\hline
\rowcolor{blue!10}
\ref{sss:comp_sep}& $\Delta P_{\rm sky} \rightarrow$ Comp sep & \multicolumn{5}{c|}{\cellcolor{gray}}& 40 & 38 & 40 & 42 & 39 & 36 & 34 & 27 & 33 & 32 \\
\hline
% Thermal on the FP
\rowcolor{blue!10}
\ref{sss:FP_heat_load}& $\Delta P_{sky} \rightarrow P_{\rm det}^{\rm cool}$ & & \multicolumn{4}{c|}{\cellcolor{gray}} & \multicolumn{5}{c|}{neg.} &  \multicolumn{5}{c|}{neg.}  \\
\rowcolor{blue!10}
& $\Delta P_{HWP} \rightarrow P_{\rm det}^{\rm cool}$ & & \multicolumn{3}{c|}{\cellcolor{gray}}& & \multicolumn{5}{c|}{neg.} &  \multicolumn{5}{c|}{neg.}  \\
\rowcolor{blue!10}
& $\Delta P_{L1} \rightarrow P_{\rm det}^{\rm cool}$ & & \multicolumn{2}{c|}{\cellcolor{gray}} & \multicolumn{2}{c|}{} & \multicolumn{5}{c|}{neg.} &  \multicolumn{5}{c|}{neg.}  \\
\rowcolor{blue!10}
& $\Delta P_{L2} \rightarrow P_{\rm det}^{\rm cool}$ & & \multicolumn{1}{c|}{\cellcolor{gray}} & \multicolumn{3}{c|}{} & \multicolumn{5}{c|}{neg.} &  \multicolumn{5}{c|}{neg.}  \\
\rowcolor{blue!10}
& $\Delta P_{T5K} \rightarrow P_{\rm det}^{\rm cool}$ & & \multicolumn{1}{c|}{\cellcolor{gray}} & \multicolumn{3}{c|}{} &  \multicolumn{5}{c|}{9} &  \multicolumn{5}{c|}{12} \\
\rowcolor{blue!10}
& $\Delta P_{T2K} \rightarrow P_{\rm det}^{\rm cool}$ & & \multicolumn{1}{c|}{\cellcolor{gray}} & \multicolumn{3}{c|}{} & \multicolumn{5}{c|}{11} &  \multicolumn{5}{c|}{12}  \\

\hline
% Thermal on the HWP
\rowcolor{blue!10}
\ref{sss:HWP_heat_load}& $\Delta P_{\rm sky} \rightarrow$ HWP & \multicolumn{4}{c|}{} & \cellcolor{gray} & \multicolumn{5}{c|}{neg.} &  \multicolumn{5}{c|}{neg.}   \\
\hline
% Thermal on the lenses
\rowcolor{blue!10}
\ref{sss:lenses_heat_load}& $\Delta P_{\rm sky} \rightarrow$ L1, L2 & \multicolumn{2}{c|}{} & \multicolumn{3}{c|}{\cellcolor{gray}} & \multicolumn{5}{c|}{neg.} &  \multicolumn{5}{c|}{neg.} \\
\hline
%-----Dynamic
\hline
%NEP stability
\rowcolor{blue!10}
\ref{sss:dynamic_NEP}& $\delta P_{\rm sky} \rightarrow $NEP & \multicolumn{5}{c|}{\cellcolor{gray}} &  \multicolumn{10}{c|}{14} \\
\hline
% FP stability
\rowcolor{blue!10}
\ref{sss:dynamic_Tdet} & $\delta P \rightarrow P_{\rm det}^{\rm cool}$ & & \multicolumn{4}{c|}{\cellcolor{gray}} &  \multicolumn{10}{c|}{neg.} \\
\hline
% Optical element stability
\rowcolor{blue!10}
\ref{sss:dynamic_Tinst}& $\delta P \rightarrow P_{\rm HWP, L1, L2}$ & \multicolumn{2}{c|}{}& \multicolumn{3}{c|}{\cellcolor{gray}} &  \multicolumn{10}{c|}{neg.} \\

\hline
\hline

%%% COB
%-----Static
\rowcolor{orange!18}
\multicolumn{17}{|c|}{ $A^{C}$ [dB]} \\
\hline
\rowcolor{orange!10}
\ref{sss:NEP}& $\Delta P \rightarrow$ NEP & \multicolumn{5}{c|}{\cellcolor{gray}} & 34 & 46 & 36 & 30 & 18 & 28 & 42 & 24 & 22 & 52 \\
\hline
\rowcolor{orange!10}
\ref{sss:comp_sep} & $\Delta P_{\rm sky} \rightarrow$ Comp sep & \multicolumn{5}{c|}{\cellcolor{gray}} & 44 & 41 & 34 & 38 & 34 & 47 & 45 & 41 & 44 & 39 \\
\hline
%-----Dynamic
\hline
\rowcolor{orange!10}
\ref{sss:dynamic_NEP} & $\delta P_{\rm sky} \rightarrow $ NEP & \multicolumn{5}{c|}{\cellcolor{gray}} &  \multicolumn{10}{c|}{35} \\

%%% HOB
\hline
\hline
\rowcolor{red!18}
\multicolumn{17}{|c|}{ $A^{H}$ [dB]} \\
\hline
%------Static
% Detection chain
\rowcolor{red!10}
\ref{sss:NEP}& $\Delta P \rightarrow $ NEP & \multicolumn{5}{c|}{\cellcolor{gray}} & 82 & 66 & 66 & 68 & 76 & 72 & 78 & 80 & 72 & 58 \\
\hline
\rowcolor{red!10}
\ref{sss:comp_sep} & $\Delta P_{sky} \rightarrow$ Comp sep & \multicolumn{5}{c|}{\cellcolor{gray}} & 69 & 66 & 64 & 60 & 69 & 66 & 64 & 60 & 65 & 62 \\
\hline
% Thermal on FP
\rowcolor{red!10}
\ref{sss:FP_heat_load}& $\Delta P_{sky} \rightarrow P_{\rm det}^{\rm cool}$ & & \multicolumn{4}{c|}{\cellcolor{gray}} & \multicolumn{5}{c|}{8} &  \multicolumn{5}{c|}{4}  \\
\rowcolor{red!10}
& $\Delta P_{HWP} \rightarrow P_{\rm det}^{\rm cool}$ & & \multicolumn{3}{c|}{\cellcolor{gray}} & & \multicolumn{5}{c|}{19} &  \multicolumn{5}{c|}{16}  \\
\rowcolor{red!10}
& $\Delta P_{L1} \rightarrow P_{\rm det}^{\rm cool}$ & & \multicolumn{2}{c|}{\cellcolor{gray}} & \multicolumn{2}{c|}{} & \multicolumn{5}{c|}{9} &  \multicolumn{5}{c|}{2}  \\
\rowcolor{red!10}
& $\Delta P_{L2} \rightarrow P_{\rm det}^{\rm cool}$ & & \multicolumn{1}{c|}{\cellcolor{gray}} & \multicolumn{3}{c|}{} & \multicolumn{5}{c|}{16} &  \multicolumn{5}{c|}{10}  \\
\rowcolor{red!10}
& $\Delta P_{T5K} \rightarrow P_{\rm det}^{\rm cool}$ & & \multicolumn{1}{c|}{\cellcolor{gray}} & \multicolumn{3}{c|}{} & \multicolumn{5}{c|}{23} &  \multicolumn{5}{c|}{17}  \\
\rowcolor{red!10}
& $\Delta P_{T2K} \rightarrow P_{\rm det}^{\rm cool}$ & & \multicolumn{1}{c|}{\cellcolor{gray}} & \multicolumn{3}{c|}{} & \multicolumn{5}{c|}{11} &  \multicolumn{5}{c|}{neg.}  \\
\hline
% Thermal on HWP
\rowcolor{red!10}
\ref{sss:HWP_heat_load}& $\Delta P_{\rm sky} \rightarrow$ HWP & \multicolumn{4}{c|}{} & \cellcolor{gray} & \multicolumn{5}{c|}{neg.} &  \multicolumn{5}{c|}{neg.} \\
\hline
% Thermal on lenses
\rowcolor{red!10}
\ref{sss:lenses_heat_load} & $\Delta P_{\rm sky} \rightarrow$ L1, L2 & \multicolumn{2}{c|}{} & \multicolumn{3}{c|}{\cellcolor{gray}} & \multicolumn{5}{c|}{neg.} & \multicolumn{5}{c|}{neg.}  \\
\hline

\hline
\hline
%------Dynamic
% NEP stability
\rowcolor{red!10}
\ref{sss:dynamic_NEP} & $\delta P_{\rm sky} \rightarrow $NEP & \multicolumn{5}{c|}{\cellcolor{gray}} &  \multicolumn{10}{c|}{15} \\
\hline
% FP stability
\rowcolor{red!10}
\ref{sss:dynamic_Tdet} & $\delta P\rightarrow P_{\rm det}^{\rm cool}$ & & \multicolumn{4}{c|}{\cellcolor{gray}} &  \multicolumn{10}{c|}{neg.} \\
\hline
% Optical element stability
\rowcolor{red!10}
\ref{sss:dynamic_Tinst} & $\delta P \rightarrow P_{\rm HWP, L1, L2}$ & \multicolumn{2}{c|}{} & \multicolumn{3}{c|}{\cellcolor{gray}} &  \multicolumn{10}{c|}{neg.} \\
\hline
\end{tabular}
\caption{Synthesis of the requirements on the attenuation factors $A^L$ (blue), $A^C$ (orange) and $A^H$ (red) in dB, derived from each analysis, either detailed per frequency channel or common for all (neg. meaning negligible compared to others analysis). The first column, titled Section, indicates which section of the article this result appears in. The column titled Position $s$ indicates in grey the combination of filter positions on which the requirement is set.}
\label{tab:summary}
\end{table}

%%%%%%%%%%%%%%%%%%%%%%%%%
%%%%%%%%%%%%%%%%%%%%%%%%%
\section{Conclusion} 
%%%%%%%%%%%%%%%%%%%%%%%%%
%%%%%%%%%%%%%%%%%%%%%%%%%
\label{sec:conclusion}

% Rappel de la méthode
In this paper, we presented a general methodology to derive specifications on out-of-band power rejection levels for a CMB instrument. The methodology was applied to the case of \textit{LiteBIRD}'s MFT and HFT telescopes. We started with high-level scientific and instrumental requirements: expected precision in the tensor-to-scalar ratio measurement and thermal constraints (available cooling power and thermal stability). Specifically, starting from the assumption of a given filtering scheme, we evaluate its impact up to the measurement of the tensor-to-scalar ratio, assuming an instrument model, a sky emission model, and a data analysis pipeline. We conducted both a static and dynamic study. In the static case, we propagated the out-of-band power up to the detection chain, exploring the impact on the detector NEP and on the component separation efficiency. We also studied the impact on the instrument thermal balance. In the dynamic case, we propagated the effect of out-of-band power fluctuations.

% Commentaires table 6
The multiple requirements derived through this work have been summarized in table~\ref{tab:summary}. We show the requirements derived from each analysis on the attenuation factors $A^L$ (blue), $A^C$ (orange) and $A^H$ (red), either detailed per frequency channel or common for all. We can outline the main points of this table. First, the highest attenuation factors required are for the HOB domain, for which we rich a maximum of 82\,dB attenuation required on $A^H$. These strict requirements stem from the study on the impact on the detection chain: an increase in the detector NEP and a less efficient component separation (section~\ref{ss:req_from_det}). Regarding the requirements derived in section~\ref{ss:req_from_thermal}, looking at the impact on the thermal balance on the instrument in the static case, we see that it is often negligible, or small compared to the requirement from the impact on the detection chain. The maximum required attenuation factor is 23\,dB, due to the thermal load exerted on the focal plane by the T5K. Finally, we observe that the requirements derived from the dynamic analysis (section~\ref{ss:req_from_dynamic}) are negligible and will not constrain the instrument's design.

% Future design
Now that we have these requirements, the next step would be to properly define the filtering scheme. While this goes beyond the scope of this paper and requires an in-depth study of existing filter technologies, we can offer some general ideas. An important conclusion, set in section~\ref{sss:HWP_heat_load}, is that an infra-red filter at position $s_4$ in front of the HWP is not needed. This is a key point, because we do not want to place any optical components upstream of the HWP in order to modulate the raw signal coming from the sky. Similarly, thanks to the lenses and the HWP which themselves act as filters, we think there is no need for filters at positions $s_3$ and $s_2$ either. By stacking several filters in position $s_1$ (above the focal plane) combined with on-chip filters in position $s_0$, we believe we can meet the requirements summarized in table~\ref{tab:summary}. According to the existing technologies -- such as metal-mesh filters~\cite{Ade_2006} -- the requirements we have identified appear to be achievable. This confidence is also based on experience gained from previously built instruments in the community, such as the Planck satellite~\cite{Ade_2010_Planck_prelaunch} and the Simons Observatory~\cite{Dicker:2018ytl}. However, this will need to be validated through simulations and, above all, laboratory measurements.

% Perspectives : adaptation au nouveau design
Finally, following the redesign of the \textit{LiteBIRD} instrument, we will adapt the method presented here to the new design. The consolidation of the three telescopes into a single reflecting one, with a new allocation of frequency bands, requires adapting the code used here. In particular, the propagation of the thermal powers inside the instrument has to be fully changed. However, we believe that the main conclusions drawn from this study will remain unchanged, although this will, of course, need to be confirmed. This work to adapt the code to the new design is already underway within the collaboration. Moreover, beyond the \textit{LiteBIRD} project, we believe this study is of interest for the community, both for current and future CMB experiments.

\appendix
%------------------------------------------------------
\section{Detector NEP calculation}
\label{app:NEP}

\begin{figure}[ht!]
    \centering
    \includegraphics[width=0.9\linewidth]{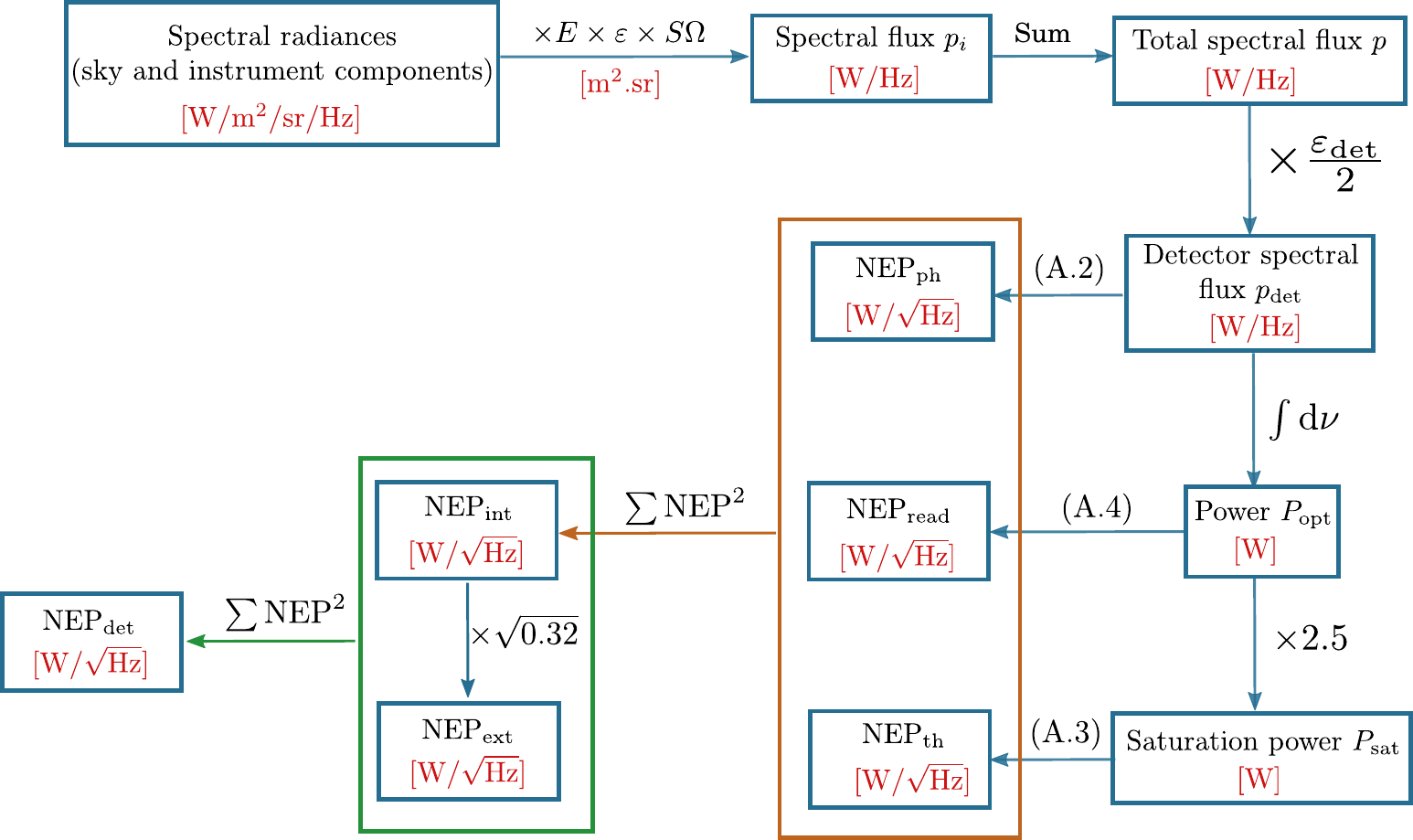}   
    \caption{Block diagram to illustrate the detector NEP calculation. Physical units are written in red.}
    \label{fig:NEP_calcul}
\end{figure}
The detector NEP calculation is sketched in figure~\ref{fig:NEP_calcul}. As seen in section~\ref{sss:NEP} we split the total detector NEP in an internal and an external contribution. The internal part includes photon NEP, thermal carrier NEP and readout NEP combined as:
\begin{equation}
    \label{eq:NEP_int}
    \rm{NEP}_{int} = \sqrt{\rm{NEP}_{ph}^2 + \rm{NEP}_{th}^2 + \rm{NEP}_{read}^2}.
\end{equation}
The photon NEP, due to the distribution of arrival times of photons at the detector input, is defined as
\begin{equation}
    \label{eq:NEP_photon}
    \rm{NEP}_{ph} = \sqrt{\int_{\nu_1}^{\nu_2}
                    \left[ 2h\nu\sum_i P_i + 2\left( \sum_i P_i\right)^2\right ] \dd\nu}
\end{equation}
where $P_i$ are the powers per unit of frequency computed in section~\ref{sss:power_det} multiplied by the detector efficiency $\varepsilon_{\rm det}$ and divided by two as \textit{LiteBIRD} TES are polarization sensitive. Note that we assume 100\,\% photon bunching and that we integrate over the bunching cross terms.
To calculate the thermal carrier NEP inherent to the bolometer power dissipation, we use the following equation
\begin{equation}
    \label{eq:NEP_thermal}
    \rm{NEP}_{th} = \sqrt{4 k_B  P_{\rm sat} T_b \frac{(n+1)^2}{2n+3}
                        \frac{(T_c/T_b)^{2n+3} - 1}{[(T_c/T_b)^{n+1} - 1]^2}}\, ,
\end{equation}
where $T_c = 0.171$\,K is the critical temperature, $T_b=0.100$\,K is the bath temperature, $n=3$ because the thermal carrier is a phonon~\cite{Mather_1982, Suzuki_2013PhD}. $P_{\rm sat} = 2.5 P_{\rm opt}$ is the detector saturation power.
Readout NEP, associated with fluctuations in the bias current across the bolometer is estimated with:
\begin{equation}
    \label{eq:NEP_read}
    \rm{NEP}_{read} = 3.5 \sqrt{2 P_{\rm opt}}\, .
\end{equation}
The sources that contribute to external NEP are various and difficult to model. This is why, we estimate the external NEP from the internal one, requiring that the total detector noise should not be impacted by external noise by more than 15\,\%. This leads to
\begin{equation}
    \label{eq:NEP_ext}
    \rm{NEP}_{ext} = \sqrt{0.32}\, \rm{NEP}_{int}.
\end{equation}
Finally, the quadratic sum of the external and internal NEP leads to the detector NEP:
\begin{equation}
    \label{eq:NEP_det}
    \rm{NEP}_{det}^2 = \rm{NEP}_{int}^2 + \rm{NEP}_{ext}^2.
\end{equation}

%------------------------------------------------------
\section{Technical details on the HWP non-ideality study}
\label{app:HWP_margin}

Technical details about the HWP non-ideality study presented in section~\ref{sss:HWP_margin} are reported here. We follow the procedure presented in \cite{Giardiello:2021uxq}, which has been ported in a \texttt{litebird\_sim} module\footnote{\url{https://github.com/litebird/litebird_sim/blob/master/litebird_sim/hwp_sys/hwp_sys.py}}. The TOD is computed as:
\begin{equation}
\begin{split} 
    d_{obs}(t_{i})\,=\,\frac{\int d\nu\,\frac{\partial BB(\nu,T)}{\partial T_{CMB}}\,\tau\left(\nu\right)\,M_{i}^{TX}(\nu)\left(m^X_{CMB}+m^X_{FG}\left(\nu\right)\right)}{\int  d\nu \frac{\partial BB(\nu,T)}{\partial T_{CMB}}\,\tau \left(\nu\right)},
\end{split}
\label{eq:dobsnu}
\end{equation}
where $\tau(\nu)$ is the bandpass with different levels of out-of-band excess, $M_{i}^{TX}(\nu)$ is the Mueller matrix of the HWP (for its derivation starting from the HWP Jones matrix, see~\cite{Giardiello:2021uxq}), $\frac{\partial BB(\nu,T)}{\partial T_{CMB}}$ is the factor converting from CMB thermodynamic units to brightness and $m^X_{CMB/FG}$ is the CMB/foreground map for the $X = T, Q, U$ field. For the foregrounds we consider the \texttt{d0s0 PySM} models.  The denominator brings the TOD back to CMB units. 

To derive a map from the TOD in Eq.~\ref{eq:dobsnu}, we perform the map-making procedure
\begin{equation}
\begin{split}    
    {m_{out}\,}&={\,\left(\sum_{i} ^{}B_{i}^{T} B_{i} \right)^{-1} \left( \sum_{i} ^{}B_{i}^{T} d_{obs}(t_{i}) \right) }, %\\
%{}&\relempty{ = \left(\sum_{i} ^{}B_{i}^{T} B_{i} \right)^{-1} \left( \sum_{i} ^{}B_{i}^{T} A_{CMB,i} \, m_{CMB}\right) + \left(\sum_{i} ^{}B_{i}^{T} B_{i} \right)^{-1}\left(\sum_{i} ^{}B_{i}^{T} \int _{}^{}A_{FG,i}(\nu) \, m_{FG}(\nu) d\nu \right)}\\
\end{split}
\label{eq:map-making}
\end{equation}
where $B_{i}$ is the map-making matrix built to recover the CMB component:
\begin{equation}
    B_{i}^{X\,=\,T,Q,U}=\,\left(\frac{\int _{}^{}d\nu \,\frac{\partial BB(\nu,T)}{\partial T_{CMB}}\,\tau _{s}\left(\nu\right)\,M_{i}^{TX}(\nu)}{\int _{}^{}d\nu \frac{\partial BB(\nu,T)}{\partial T_{CMB}}\,\tau _{s}\left(\nu\right)}\right).
\end{equation}

In the map-making procedure we adopt an ideal, top-hat bandpass $\tau_s(\nu)$ with no out-of-band excess, to represent our ignorance of the level of out-of-band contamination. The HWP Mueller matrix $M_{i}^{TX}$ is the same used in the TOD, not to introduce any mismatch in our estimate of the HWP parameters. 

To derive the residual power due to the out-of-band contamination, we build a template map $m_{\rm templ}$ with the same procedure we have just outlined, but using the same ideal, top-hat bandpass profile both in the TOD and in the map-making, $\tau(\nu) = \tau_s(\nu)$. The residual map is computed as $m_{\rm res} = m_{\rm out} - m_{\rm templ}$, for each frequency channel and each level of out-of-band excess. We refer to~\cite{Giardiello:2021uxq} for more details on the map-making procedure and the derivation of residual maps.

%------------------------------------------------------
\section{Solid angle calculation}
\label{app:solid_angle}

\begin{figure}
    \label{fig:sketch_solid_angle}
    \centering
    \includegraphics[width=0.6\linewidth]{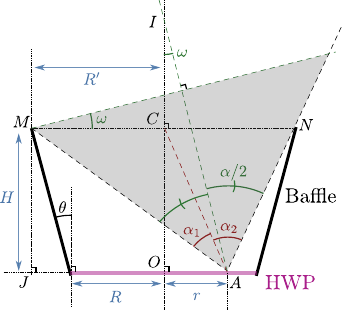}
    \caption{Sketch of the telescope upper part, HWP and baffle. We consider a point $A$ belonging to the HWP, located at a distance $r$ from the optical axis (OC). The grey zone shows the solid angle under which point $A$ sees the sky.}
\end{figure}

We present here the calculation to get the solid angle expression given in equation~\eqref{eq:solid_angle}. The notations are defined in figure~\ref{fig:sketch_solid_angle}.

The solid angle under which a point $A$ belonging to the HWP, located at a distance $r$ from the optical axis, sees the sky is computed using spherical coordinates as
\begin{equation}
    \label{eq:solid_angle}
    \Omega(r) = \int_0^{2\pi}\int_0^{\alpha(r)/2} \cos(\omega(r))\sin\theta\dd\theta\dd\varphi
\end{equation}
where the angles $\alpha(r)$ and $\omega(r)$ are defined in figure~\ref{fig:sketch_solid_angle}. This expression can be simplified and expressed as a function of known quantities: the baffle height $H$, the baffle aperture angle $\theta$ and the HWP radius $R$. Then, once we have the solid angle as a function of $r$, the optical extent is obtained by integrating over the HWP surface.

The detailed calculation is done as follow. We consider a point $A$ belonging to the HWP, located at a distance $r$ from the centre. We know the radius $R$ of the HWP, the angle of the baffle $\theta=14\deg$ and the baffle height $H$. From those quantities, we first estimate the angle $\alpha(r) = \alpha_1(r) + \alpha_2(r)$. Using the Al-Kashi theorem in ACM triangle, we have
\begin{equation}
    R'^2 = AM^2 + AC^2 - 2 AM.AC\cos\alpha_1
\end{equation}
where $R' = R + H\tan\theta$ is the baffle radius at the top. Moreover, thanks to Pythagore theorem, we have $AC^2 = r^2 + H^2$ and $AM^2 = H^2 + (R' + r)^2$.
Thus, by combining these results, we obtain
\begin{equation}
    \cos\alpha_1 = \frac{H^2 + rR' + r^2}{\sqrt{(H^2 + (R' + r)^2)(r^2 + H^2)}}.
\end{equation}
In the same way, one can show that
\begin{equation}
    \cos\alpha_2 = \frac{H^2 - rR' + r^2}{\sqrt{(H^2 + (R' - r)^2)(r^2 + H^2)}}.
\end{equation}

Once we have $\alpha(r)$, we can compute the solid angle $\Omega(r)$. We have $\dd\Omega(r) = \cos(\omega(r))\sin\theta\dd\theta\dd\phi$ and $\cos(\omega(r)) = \sin\left(\beta + \frac{\alpha(r)}{2}\right)$ with $\tan\beta = \frac{H}{R' + r}$. 

Thus, we can express the solid angle $\Omega(r)$ as a function of known quantities only:
\begin{align}
    \Omega(r) &= \int_0^{2\pi}\int_0^{\alpha(r)/2} \cos(\omega(r))\sin\theta\dd\theta\dd\varphi\\
    &= 2\pi \sin{\left[\arctan\left(\frac{H}{R'+r}\right) + \frac{\alpha(r)}{2} \right]} \left(1 - \cos\left(\frac{\alpha(r)}{2}\right) \right).
\end{align}

To get the solid angle under which point $A$ sees the baffle, we simply make the difference between the half-sphere solid angle $2\pi$ and $\Omega(r)$. Others solid angles are computed similarly.

Finally, in the case of sky power, we assume conservation of the optical extent along the system in the far field regime. Thus, for the sky seen by L1, L2 or the focal plane, we expressed the optical extent as $S\Omega = S \pi\theta^2$ where $S$ is the HWP surface and $\theta$ the baffle aperture angle.\footnote{Here the solid angle expression is simplified in the small angle approximation.}

\newpage
%------------------------------------------------------
\section{Power detailed calculation}
\label{app:det_power}

\begin{table}[ht!]
\centering
\tiny
\begin{tabular}{|c|c|c|c|c|c||c|c|c|c|c|}
\hline
& \multicolumn{5}{c||}{\textbf{MFT}} & \multicolumn{5}{c|}{\textbf{HFT}} \\
\hline
\textbf{Channel $i$ [GHz]} & 100 & 119 & 140 & 166 & 195 & 195 & 235 & 280 & 337 & 402 \\
\hline
\textbf{$\widetilde P^L_{c,i}$ [pW]} & \multicolumn{10}{c|}{} \\
\hline
CMB & 9.01e-01 & 7.12e-01 & 5.98e-01 & 5.27e-01 & 4.94e-01 & 8.04e-01 & 5.99e-01 & 4.97e-01 & 4.29e-01 & 3.15e-01 \\
Gal & 3.22e-03 & 2.81e-03 & 2.55e-03 & 2.37e-03 & 2.28e-03 & 1.60e-02 & 1.35e-02 & 1.20e-02 & 1.09e-02 & 8.86e-03 \\
Syn & 8.19e-04 & 6.72e-04 & 5.78e-04 & 5.17e-04 & 4.89e-04 & 2.34e-04 & 1.80e-04 & 1.51e-04 & 1.32e-04 & 9.83e-05 \\
OB & 6.03e-06 & 5.13e-06 & 4.77e-06 & 4.53e-06 & 4.41e-06 & 1.27e-05 & 1.11e-05 & 1.02e-05 & 9.41e-06 & 7.87e-06 \\
IPD & 2.85e-04 & 2.47e-04 & 2.21e-04 & 2.02e-04 & 1.93e-04 & 5.60e-04 & 4.52e-04 & 3.91e-04 & 3.47e-04 & 2.68e-04 \\
HWP & 3.31e-01 & 2.62e-01 & 2.20e-01 & 1.94e-01 & 1.82e-01 & 8.30e-01 & 6.19e-01 & 5.14e-01 & 4.43e-01 & 3.25e-01 \\
L1 & 7.86e-02 & 6.21e-02 & 5.22e-02 & 4.59e-02 & 4.31e-02 & 1.13e-01 & 8.43e-02 & 7.00e-02 & 6.04e-02 & 4.43e-02 \\
L2 & 8.38e-02 & 6.62e-02 & 5.56e-02 & 4.90e-02 & 4.60e-02 & 1.14e-01 & 8.54e-02 & 7.09e-02 & 6.11e-02 & 4.48e-02 \\
Apt & 1.83e-01 & 1.44e-01 & 1.21e-01 & 1.07e-01 & 1.00e-01 & 2.59e-01 & 1.93e-01 & 1.60e-01 & 1.38e-01 & 1.01e-01 \\
T5K & 1.56e-01 & 1.23e-01 & 1.04e-01 & 9.14e-02 & 8.58e-02 & 1.01e-01 & 7.55e-02 & 6.27e-02 & 5.41e-02 & 3.97e-02 \\
T2K & 3.68e-02 & 2.91e-02 & 2.45e-02 & 2.15e-02 & 2.02e-02 & 7.75e-03 & 5.78e-03 & 4.79e-03 & 4.14e-03 & 3.03e-03 \\
Ref1 & 7.92e-03 & 1.79e-03 & 6.15e-04 & 1.36e-03 & 2.38e-03 & 1.47e-02 & 2.44e-03 & 2.88e-04 & 1.65e-03 & 3.59e-03 \\
\hline
Total& 1.78 & 1.4 & 1.18 & 1.04 & 0.98 & 2.26 & 1.68 & 1.39 & 1.2 & 0.89 \\
\hline
\hline
\textbf{$\widetilde P^D_{c,i}$ [pW]} & \multicolumn{10}{c|}{} \\
\hline
CMB & 4.78e-01 & 6.91e-01 & 7.66e-01 & 7.46e-01 & 7.73e-01 & 3.88e-01 & 3.10e-01 & 2.27e-01 & 1.38e-01 & 4.65e-02 \\
Gal & 4.52e-03 & 1.16e-02 & 2.22e-02 & 3.92e-02 & 7.55e-02 & 3.76e-02 & 7.18e-02 & 1.27e-01 & 2.24e-01 & 2.64e-01 \\
Syn & 1.65e-05 & 1.78e-05 & 1.56e-05 & 1.23e-05 & 1.11e-05 & 5.54e-06 & 4.26e-06 & 3.35e-06 & 2.54e-06 & 1.29e-06 \\
OB & 4.72e-06 & 9.17e-06 & 1.44e-05 & 2.08e-05 & 3.26e-05 & 1.61e-05 & 2.55e-05 & 3.76e-05 & 5.47e-05 & 5.56e-05 \\
IPD & 2.23e-04 & 4.41e-04 & 6.65e-04 & 9.23e-04 & 1.42e-03 & 7.09e-04 & 1.03e-03 & 1.43e-03 & 1.99e-03 & 1.86e-03 \\
HWP & 2.61e-01 & 3.19e-01 & 3.57e-01 & 4.16e-01 & 6.07e-01 & 8.85e-01 & 7.33e-01 & 7.37e-01 & 9.06e-01 & 8.55e-01 \\
L1 & 6.62e-02 & 1.07e-01 & 1.43e-01 & 1.85e-01 & 2.49e-01 & 1.11e-01 & 1.26e-01 & 1.49e-01 & 1.72e-01 & 1.17e-01 \\
L2 & 7.12e-02 & 1.14e-01 & 1.52e-01 & 1.98e-01 & 2.66e-01 & 1.13e-01 & 1.24e-01 & 1.44e-01 & 1.64e-01 & 1.11e-01 \\
Apt & 1.16e-01 & 9.98e-02 & 1.06e-01 & 1.21e-01 & 1.52e-01 & 1.84e-01 & 8.94e-02 & 8.09e-02 & 9.19e-02 & 5.43e-02 \\
T5K & 1.29e-01 & 1.94e-01 & 2.39e-01 & 2.76e-01 & 3.52e-01 & 1.10e-01 & 1.48e-01 & 2.02e-01 & 2.34e-01 & 1.42e-01 \\
T2K & 2.08e-02 & 2.53e-02 & 2.48e-02 & 2.16e-02 & 1.96e-02 & 3.12e-03 & 2.61e-03 & 2.01e-03 & 1.17e-03 & 2.94e-04 \\
Ref1 & 6.19e-03 & 2.32e-03 & 6.49e-04 & 2.63e-03 & 7.15e-03 & 1.37e-02 & 2.84e-03 & 7.44e-05 & 2.36e-03 & 4.85e-03 \\
\hline
Total& 1.15 & 1.57 & 1.81 & 2.01 & 2.5 & 1.85 & 1.61 & 1.67 & 1.94 & 1.6 \\
\hline
\hline
\textbf{$\widetilde P^C_{c,i}$ [pW]} & \multicolumn{10}{c|}{} \\
\hline
CMB & 2.84e+00 & 1.93e+00 & 1.44e+00 & 1.20e+00 & 1.05e+00 & 6.47e-01 & 4.63e-01 & 4.14e-01 & 4.15e-01 & 3.59e-01 \\
Gal & 1.41e-01 & 1.16e-01 & 9.30e-02 & 6.79e-02 & 2.77e-02 & 7.74e-01 & 6.11e-01 & 4.81e-01 & 3.30e-01 & 1.85e-01 \\
Syn & 4.67e-05 & 3.41e-05 & 2.91e-05 & 2.76e-05 & 2.66e-05 & 1.08e-05 & 8.27e-06 & 7.20e-06 & 6.65e-06 & 5.56e-06 \\
OB & 6.70e-05 & 5.19e-05 & 4.23e-05 & 3.31e-05 & 1.99e-05 & 1.66e-04 & 1.35e-04 & 1.09e-04 & 8.10e-05 & 5.79e-05 \\
IPD & 3.15e-03 & 2.49e-03 & 1.95e-03 & 1.47e-03 & 8.69e-04 & 7.27e-03 & 5.41e-03 & 4.14e-03 & 2.95e-03 & 1.95e-03 \\
HWP & 1.79e+00 & 1.30e+00 & 1.00e+00 & 7.81e-01 & 5.17e-01 & 4.11e+00 & 2.99e+00 & 2.35e+00 & 1.76e+00 & 1.10e+00 \\
L1 & 7.19e-01 & 5.13e-01 & 3.79e-01 & 2.74e-01 & 1.82e-01 & 6.68e-01 & 4.56e-01 & 3.33e-01 & 2.44e-01 & 1.89e-01 \\
L2 & 7.67e-01 & 5.49e-01 & 4.05e-01 & 2.92e-01 & 1.94e-01 & 6.41e-01 & 4.38e-01 & 3.22e-01 & 2.39e-01 & 1.85e-01 \\
Apt & 4.83e-01 & 3.74e-01 & 2.92e-01 & 2.30e-01 & 1.76e-01 & 3.53e-01 & 3.11e-01 & 2.51e-01 & 1.95e-01 & 1.56e-01 \\
T5K & 1.09e+00 & 7.67e-01 & 5.69e-01 & 4.35e-01 & 3.16e-01 & 8.64e-01 & 5.78e-01 & 4.00e-01 & 2.86e-01 & 2.40e-01 \\
T2K & 8.26e-02 & 5.64e-02 & 4.39e-02 & 3.88e-02 & 3.71e-02 & 5.22e-03 & 3.61e-03 & 3.16e-03 & 3.28e-03 & 2.97e-03 \\
Ref1 & 5.26e-02 & 8.55e-03 & 1.16e-03 & 4.24e-03 & 7.02e-03 & 5.76e-02 & 8.03e-03 & 1.70e-04 & 4.04e-03 & 1.10e-02 \\
\hline
Total& 7.97 & 5.62 & 4.23 & 3.32 & 2.51 & 8.13 & 5.87 & 4.57 & 3.48 & 2.43 \\
\hline
\hline
\textbf{$\widetilde P^H_{c,i}$ [pW]} & \multicolumn{10}{c|}{} \\
\hline
CMB & 4.96e-01 & 3.92e-01 & 3.29e-01 & 2.90e-01 & 2.72e-01 & 1.28e-02 & 9.58e-03 & 7.95e-03 & 6.86e-03 & 5.03e-03 \\
Gal & 1.44e+02 & 1.26e+02 & 1.14e+02 & 1.06e+02 & 1.02e+02 & 4.63e+01 & 3.89e+01 & 3.47e+01 & 3.16e+01 & 2.56e+01 \\
Syn & 4.40e-05 & 3.62e-05 & 3.11e-05 & 2.78e-05 & 2.63e-05 & 1.02e-05 & 7.83e-06 & 6.60e-06 & 5.75e-06 & 4.29e-06 \\
OB & 2.13e+02 & 1.81e+02 & 1.68e+02 & 1.60e+02 & 1.56e+02 & 7.05e+01 & 6.19e+01 & 5.66e+01 & 5.24e+01 & 4.38e+01 \\
IPD & 2.73e+02 & 2.37e+02 & 2.11e+02 & 1.94e+02 & 1.85e+02 & 8.48e+01 & 6.84e+01 & 5.92e+01 & 5.25e+01 & 4.05e+01 \\
HWP & 2.98e+03 & 2.35e+03 & 1.98e+03 & 1.74e+03 & 1.63e+03 & 7.20e+02 & 5.37e+02 & 4.46e+02 & 3.85e+02 & 2.82e+02 \\
L1 & 2.15e+01 & 1.70e+01 & 1.43e+01 & 1.26e+01 & 1.18e+01 & 2.43e+00 & 1.81e+00 & 1.50e+00 & 1.30e+00 & 9.51e-01 \\
L2 & 4.31e+01 & 3.40e+01 & 2.86e+01 & 2.52e+01 & 2.36e+01 & 4.85e+00 & 3.62e+00 & 3.00e+00 & 2.59e+00 & 1.90e+00 \\
Apt & 2.12e+00 & 1.67e+00 & 1.41e+00 & 1.24e+00 & 1.16e+00 & 2.85e-01 & 2.13e-01 & 1.77e-01 & 1.52e-01 & 1.12e-01 \\
T5K & 4.89e+00 & 3.86e+00 & 3.25e+00 & 2.86e+00 & 2.68e+00 & 7.46e-01 & 5.56e-01 & 4.61e-01 & 3.98e-01 & 2.92e-01 \\
T2K & 3.39e-02 & 2.68e-02 & 2.25e-02 & 1.98e-02 & 1.86e-02 & 1.87e-04 & 1.39e-04 & 1.16e-04 & 9.99e-05 & 7.32e-05 \\
Ref1 & 2.30e-01 & 3.70e-02 & 7.46e-04 & 1.94e-02 & 4.74e-02 & 4.57e-02 & 6.87e-03 & 3.89e-05 & 3.93e-03 & 9.97e-03 \\
\hline
Total& 3680.31 & 2953.89 & 2518.79 & 2242.4 & 2116.63 & 930.37 & 712.82 & 601.48 & 525.68 & 395.44 \\
\hline
\hline
\end{tabular}
\caption{Incident $\widetilde P^d_{c,i}$ power in pW falling on a detector for the ten MFT and HFT frequency channels $i$, in the four frequency domains: LOB, DIB, COB and HOB and for each sky and instrument component.}
\label{tab:total_power}
\end{table}

\acknowledgments
This work was supported by a CNES postdoctoral fellowship. We thank the internal reviewer of the \textit{LiteBIRD} collaboration E. Battistelli and M. Calvo who helped a lot in improving the paper as well as J. Macias Perez who coordinated the review. Finally, we would like to thank the editor and the JCAP reviewer, whose input greatly improved the article.

\newpage
% Biblio
\bibliographystyle{JHEP}
\bibliography{Papier_LB_OOB} 

\end{document}